\pdfoutput=1
\documentclass[sn-basic]{sn-jnl}

\usepackage{graphicx}%
\usepackage{xr-hyper}
\usepackage{multirow}%
\usepackage{amsmath,amssymb,amsfonts}%
\usepackage{amsthm}%
\usepackage{mathrsfs}%
\usepackage[title]{appendix}%
\usepackage{xcolor}%
\usepackage{textcomp}%
\usepackage{manyfoot}%
\usepackage{booktabs}%
\usepackage{algorithm}%
\usepackage{algorithmicx}%
\usepackage{algpseudocode}%
\usepackage{listings}%

\usepackage{hyperref}
\graphicspath{ {./images/} }

\makeatletter
\newcommand*{\addFileDependency}[1]{
\typeout{(#1)}
\@addtofilelist{#1}
\IfFileExists{#1}{}{\typeout{No file #1.}}
}\makeatother

\newcommand*{\myexternaldocument}[1]{%
\externaldocument{#1}%
\addFileDependency{#1.tex}%
\addFileDependency{#1.aux}%
}

\myexternaldocument{sn-supplementary}

\begin{document}

\title[Does Transport Inequality Perpetuate Housing Insecurity?]{Does Transport Inequality Perpetuate Housing Insecurity?}


\author*[1]{\fnm{Nandini} \sur{Iyer}}\email{niyer@biocomplexlab.org}

\author[1,2]{\fnm{Ronaldo} \sur{Menezes}}\email{R.Menezes@exeter.ac.uk}

\author[1]{\fnm{Hugo} \sur{Barbosa}}\email{h.barbosa@exeter.ac.uk}

\affil*[1]{\orgdiv{BioComplex Laboratory, Computer Science}, \orgname{University of Exeter}, \orgaddress{\city{Exeter}, \country{UK}}}

\affil[2]{\orgdiv{Computer Science}, \orgname{Federal University of Cear\'a}, \orgaddress{\city{Fortaleza},  \country{Brazil}}}


\abstract{With trends of urbanisation on the rise, providing adequate housing to individuals remains a complex issue to be addressed. Often, the slow output of relevant housing policies, coupled with quickly increasing housing costs, leaves individuals with the burden of finding housing that is affordable and safe. In this paper, we unveil how urban planning, not just housing policies, can prevent individuals from accessing better housing conditions. We begin by proposing a clustering approach to characterising levels of housing insecurity in a city, by considering multiple dimensions of housing. Then we define levels of transit efficiency in 20 US cities by comparing public transit journeys to car-based journeys. Finally, we use geospatial autocorrelation to highlight how commuting to areas associated with better housing conditions results in transit commute times of over 30 minutes in most cities, and commute times of over an hour in some cases. Ultimately, we show the role that public transportation plays in locking vulnerable demographics into a cycle of poverty, thus motivating a more holistic approach to addressing housing insecurity that extends beyond changing housing policies.}




\maketitle

\section{Introduction}\label{sec1}

The rapidly increasing density of urban areas threatens to exacerbate the ever present housing crises around the world \cite{yao2014housing, goldblum2000growth}. 
Housing crises are often linked to economic and political forces, as seen by the Global Financial Crisis of 2007 \cite{clark2013aftermath, jones2014housing, wetzstein2017global}. However, Marcuse and Madden suggest that the term `crisis' wrongly implies that negative consequences of housing are not intentional \cite{marcuse2016defense}. This idea was more recently supported by \cite{heslop2020politics}, 
who indicate that the term `crisis' was invoked to justify deregulation policies under the veil of increasing housing accessibility. These examples serve, not as a political critique, but rather as an illustration of the role policies play in shaping the housing landscape. In many cases, the proposed solution by government officials and urban planners, alike, is to build more housing, particularly more affordable housing \cite{molloy2020effect, edwards2016housing}. These policies influence the state of housing in a direct and indirect manner. Direct policies range from zoning laws to rental-assistance voucher programs, and are dictated by individuals in power \cite{von2012history, antievictionMP}. On the other hand, indirect policies include the strategic planning of employment opportunities and the robustness of transport systems \cite{yates2011housing}, which can affect the level of investment that goes into developing infrastructure for different regions. Our work focuses on the effect of these indirect policies on housing insecurity, which estimates the severity of housing conditions. Specifically, we analyse how transport, employment, and urban layouts may contribute to the hardships endured by those facing housing insecurity.



There is a general consensus among researchers in transport policy and human mobility that residential and employment locations largely determine how individuals interact with their urban environment \cite{hickman2010planning,bartel1979migration}. In order to develop a deeper, more accurate, understanding of the factors that shape an urban experience, it is, therefore, essential to explore the dynamics of housing, employment and mobility. Recently, there has been an increased push to improve the quality of housing insecurity data \cite{desmond2012eviction, tourangeau2014hard}. This movement has opened the gates to facilitate various research on how housing insecurity is encoded within the urban lifestyle \cite{soyinka2018urban,roy2012urban}. Research in transport disadvantage and social exclusion illustrate how vulnerable populations face mobility constraints due to a lack of reliable transportation services from their residential locations \cite{lucas2012transport}. The field of transport poverty is still fairly nascent. As a consequence, many definitions exist to characterise transport poverty, ranging from rates of (forced) car ownership to measures of how urban systems aid accessibility \cite{lucas2016transport,mattioli2017forced}. On the other hand, there are a bulk of studies that take a more quantitative approach and use mobility data to estimate how individuals navigate between their residential and employment locations. One such study uses transit smart card data to infer home and workplace locations to ultimately categorise city dwellers into four mobility groups depending on how frequently they switch residences and jobs \cite{huang2018tracking}. A separate body of work analyses how transport facilitates commutes across subsidised housing and employment opportunities, finding correlations between high public transit accessibility and maintaining employment \cite{blumenberg2014driving}. 

All of these studies identify mobility disadvantages among the urban poor, but tend to neglect the role that urban stressors and processes, such as housing insecurity, play in creating these disadvantages. 
This oversight is prevalent both in sociological housing research and mobility studies. Research in urban mobility accounts for demographic-dependent travel behaviour, but neglects to address how residential instability may obscure mobility metrics that capture disparities across different populations. A small subset of quantitative research has combined housing and urban mobility data. Their research builds upon important and fundamental studies that identify relationships between housing characteristics and the sociodemographic composition of an urban area. However, it extends beyond that by analysing housing insecurity in conjunction with fluctuating urban processes. For instance, three distinct relocation patterns were identified in Shenzhen, China by combining transit smart card data with metrics of housing affordability \cite{gao2018exploring}. Specifically, a corridor-relocation pattern emerged, where city dwellers moved from inner urban areas to less central locations. These patterns shine light on transit constraints, as the relocation trajectories were denser along transit lines. One qualitative study observed higher rates of transit-dependency for individuals that qualify for affordable housing \cite{bardaka2019comparing}. A separate body of work takes another stride in addressing the gap between housing insecurity and mobility studies by assessing the relationship between housing affordability and car dependence. Cao and Hickman develop an index based on oil and housing prices to identify regions in London that have low levels of housing affordability, yet high levels of car dependency \cite{cao2018car}. Our work aims to continue the trajectory of these works by incorporating the dynamic nature of housing insecurity into urban analytics.

We begin by introducing a multidimensional approach to estimating housing insecurity, which we apply to 20 US cities to identify neighbourhoods that are particularly vulnerable to housing insecurity. Then, we leverage public transit networks in each of the 20 cities to define the efficiency of transport, when compared to driving times. We compare cities' transit systems' at a national scale while also considering the efficiency of transport as a function of distance, within each city. Finally, we incorporate employment data to explore the relationship between individuals' residential and employment areas. We consider how commuting times would be affected if individuals in vulnerable neighbourhoods were to work in employment areas that employ the more privileged housing demographics. Given empirical commuting behaviour, we observe average commute times of less than a 30 minutes in all 20 cities, when using cars as a mode of transport. On the other hand, commuting via public transit increases the average commute time to over 30 minutes for the majority of the cities analysed. We find that seeking better employment opportunities, while also depending on transport as a mode of commuting, results in increased commuting time for most cities, with travel times of more than an hour for 10 of the 20 cities. Ultimately, we identify disparities in the efficiency with which public transit serves different types of employment hubs. Thus, by exploring how transportation networks connect housing and employment landscapes, we underscore how transport infrastructure can create hurdles for individuals to break the cycle of housing insecurity.

\section{Data \label{sec:data}}

In order to asses how transport and employment inequalities pose additional burdens to individuals facing housing insecurity, we draw upon secondary data sources to define the state of housing, the employment landscape, and transit systems for various cities in the US.

\subsection{Housing Data Sources \label{ssec:housing_data}}
In short, housing insecurity can be distilled into seven categories: Housing Stability, Housing Affordability, Housing Quality, Housing Safety, Neighbourhood Safety, Neighbourhood Quality, and Homelessness. In Section \ref{ssec:housing_insecurity} of the Supplementary Materials, we describe characteristics of these seven housing dimensions as defined by \cite{cox2019road}. Cox points out how many quantitative studies tend to use one dimension as a proxy for housing insecurity, and therefore only capture particular disadvantages. Accordingly, in this work, we attempt to define housing insecurity using multiple dimensions. However, we do not include the Neighbourhood Safety, Neighbourhood Quality, and Homelessness dimensions as the available data sources provide information at larger geographical units. Thus, incorporating these dimensions would require sacrificing the census tract granularity at which we measure housing insecurity. Cox states that the Homelessness dimension is optional in defining housing insecurity, bolstering the decision to not include it in our definition. 
Finally, we combine the Housing Quality and Safety dimensions because their data sources largely overlapped. 

The majority of our data is sourced from the 2019 American Community Survey (ACS). This enables us to apply our analysis to various cities within the USA. We define housing characteristics at the census tract level. Census tracts are subdivisions of a county and aim to have a population of 4,000, although the population can range from 1,200 and 8,000 people. Although the ACS provides housing data at a census block group level, which are statistical divisions of census tracts, the data availability of eviction rates is limited to census tract scale. Thus, the data resolution is limited to the tract level. We measure Housing Affordability using the fraction of each tract that is severely rent burdened (spending 50\% or more of their income on housing), the median mortgage status, and the number of housing units per capita. The fraction of housing units in each tract that have complete plumbing facilities, kitchen appliances, and telephone service inform the level of Housing Quality for each census tract. Finally, Housing Stability is defined using eviction rates and levels of overcrowding within each tract. All of the listed data, barring eviction rates, is defined by the ACS. The Eviction Lab \cite{evictionlabdata} provides rates of tract-level evictions for various cities in the USA. Section \ref{sssec:housing_stability} in the Supplementary Materials discusses issues in the collection and availability of eviction data. A more detailed description of the ACS tables that are used and the data preparation are outlined in Section \ref{sec:si_data_souurces} of the Supplementary Materials.

\subsection{Public Transportation Data Sources \label{ssec:transit_data}}

The public transportation data that we use is from The Mobility Database \cite{mobilitydatabase}, which provides a means for extracting GTFS feeds for specific cities. These provide information on stops, schedules, and routes for different forms of public transportation ranging from buses to ferries. 
In this manner, we can create a workflow to develop multi-modal transit networks for different cities.  We use UrbanAccess, an open-source tool provided by the Urban Data Science Toolkit, to interpret the transit feeds data for various cities  \cite{blanchard2017urbanaccess}. This tool builds a transit-pedestrian network by combining transit networks, created using the aforementioned GTFS feeds, and pedestrian networks. Each node in the transit network represents a transit stop and edges capture successive stops on transit lines, capturing the minutes of travel between adjacent stops. These edges are weighted using data from the GTFS transit schedules. If data for a particular stop is missing, it is predicted using linear interpolation. The pedestrian network is built using OpenStreetMap. By specifying a bounding box, UrbanAccess leverages OSM's compatibility with NetworkX to build a network where nodes are particular points in the region and edges represent linear paths such as roads. Urban Access then merges these two networks to create a more comprehensive travel network. This is done by joining each pedestrian node to its closest transit node. The weight of that edge reflects the time it takes to walk from a pedestrian node to a transit node (or vice versa). The walking time is calculated using the distance between nodes and assuming a walking speed of 3 miles per hour. This network is fed into Pandana, a Python library for calculating accessibility metrics. 

\subsection{Employment Data Sources \label{ssec:employment_data}}
Given that employment opportunities greatly influence residential choices \cite{hickman2010planning}. Households that are facing forced moves may be limited in housing choices due to constraints in job opportunities. Thus, we incorporate employment data from the Longitudinal Employer-Household Dynamics program (LEHD), to understand commuting flows across census tracts. We use the LEHD Origin-Destination Employment Statistics (LODES), which is available across each state in the United States on a census tract level. LODES provides characteristics of survey participants with respect to the census tract that they live in and the census tract in which they work. This information includes income groups, industrial sectors, educational attainment, sex, race, and age. By combining a census tract's housing and public transit characteristics with its employment attributes, we can explore 
how individuals from various housing demographics may have access to different types and magnitudes of employment opportunities.

\section{Exploring Measures of Housing Insecurity \label{sec:housing}}
In this section, we detail the multidimensional approach to defining census tracts based on their level of vulnerability to housing insecurity. We apply spectral clustering to the data introduced in Section \ref{ssec:housing_data} and then evaluate the validity of the identified clusters using a range of sociodemographic indicators.

\subsection{Defining Vulnerable Housing Regions: A Clustering Approach \label{ssec:defining_vuln_housing}}
In order to examine the state of housing in various US cities, we adopt an unsupervised learning approach to define three different housing categories: most vulnerable, mildly vulnerable, and less vulnerable. We apply spectral clustering on the housing features that we outline in Section \ref{ssec:housing_data}. Details about these features are summarised in Figure \ref{fig:housing_data_summary} of the Supplementary Materials. Spectral clustering is a particularly useful approach for clustering high dimensional data, as it makes no assumptions about the shapes of clusters. Alternative approaches for clustering such as  Expectation-Maximization and K-Means are extremely sensitive to initialization. The process of spectral clustering can be broken into three steps. First, one must extract an affinity matrix from a graph that is built using the data points. The next step is spectral embedding, which leverages properties of the Graph Laplacian to represent data points in a low-dimensional space. The last step of the algorithm involves applying a classical clustering algorithm, typically K-means, to partition the embedded data into respective clusters. More details about how we preprocess housing data can be found in Section \ref{sec:si_data_prep_and_clustering} of the Supplementary Materials.

\begin{figure}[H]
  \centering
  \includegraphics[width=1 \textwidth]{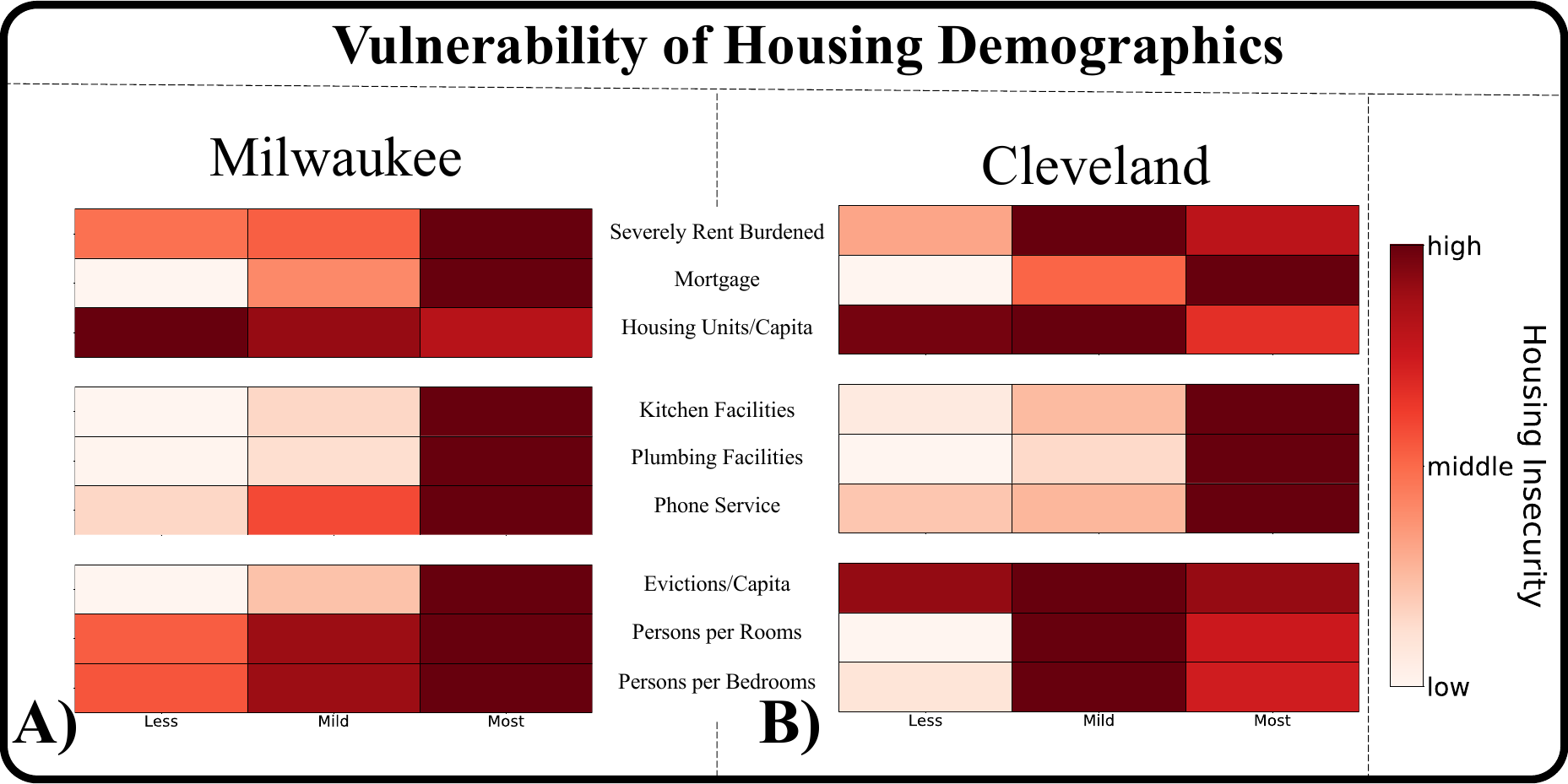}
  \caption{Heat map visualising the average housing characteristics for each housing demographic. From top to bottom, each heat map captures housing insecurity levels for the dimensions of affordability, safety/quality, and stability, respectively. Columns are the housing demographics, while rows are the housing features. Darker hues of red indicate higher levels of housing insecurity. On one hand, Milwaukee (Panel A) reflects a case where the most vulnerable census tracts consistently have the highest rates of housing insecurity. On the other hand, Cleveland (Panel B) presents a convincing case against using a single housing feature as a proxy for housing insecurity.}
  \label{fig:housing_features}
\end{figure}

Since we are considering a variety of urban areas, the number of clusters that are appropriate for each city's housing characteristics differs. To extract meaning from each cluster, we rank the housing clusters based on the mean values of their housing features, where larger values denote worse housing conditions. The group that has higher ranks across the housing dimensions is deemed most vulnerable to housing insecurity. Then, to address the varying number of clusters across cities, we partition the ranked clusters into three housing demographic groups, in which each group contains a similar number of clusters. These final three groups reflect the tracts that are the most vulnerable, mildly vulnerable, and less vulnerable to housing insecurity, in the context of each city. Figure \ref{fig:housing_features} illustrates the housing features for the resulting housing demographics in Milwaukee (Panel A) and Cleveland (Panel B). Each column represents the final housing demographic groups (Less, Mildly, and Most Vulnerable), while each row illustrates the housing characteristics that were used to define the housing demographic with respect to a city. In this manner, each cell can be defined by the following equation:
\begin{equation}\label{eq:housing_feat}
    M_{f,h} = \frac{1}{|T_h|}\sum_{t}^{T_h}\textit{HC} (f,t)
\end{equation} 
where $\textit{HC}(f,t)$ refers to the value of a housing feature, $f$, for a tract, $t$. Accordingly, for a given housing feature (row), $f$, and housing demographic (column), $h$, a cell's value is defined by averaging the housing feature for each census tract in $h$. Then, we apply row-wise normalisation to compare the differences in demographics within each city. The upper left cell of a heat map, for example, conveys the mean percentage of severely rent burdened households across all census tracts in the less vulnerable housing cluster. 

Panel A in Figure \ref{fig:housing_features} uses Milwaukee as an example for when housing demographics are clearly distinguishable, in terms of having consistent levels of housing insecurity across most features. In these types of cities, using a single housing feature as a proxy for housing insecurity could be an adequate estimation. However, the housing characteristics in Cleveland emphasise the need for a multidimensional approach to defining housing insecurity, illustrating how neighbourhoods may be vulnerable to various forms of housing insecurity, ultimately underscoring the complexities of housing conditions. For instance, the less vulnerable census tracts in Cleveland have higher insecurity than the most vulnerable tracts, in terms of  housing stock within the city. Moreover, the mildly vulnerable tracts have the highest insecurity when considering rent burden, housing stock, and the housing stability dimension, represented by the bottom-most group of heat maps. When considering each housing feature, however, the most vulnerable tracts have the highest levels of insecurity overall. The intricacies of housing conditions, then, becomes clear, with Figure \ref{fig:housing_features} emphasising the importance of considering the multidimensional nature of housing.


We validate the classification in Section \ref{ssec:housing_demographics} of the Supplementary Materials against other socioeconomic indicators of inequality.  of the 20 cities we show

Ultimately, this section highlights the pertinence of understanding the dynamics of residential mobility, which influences the accuracy of inequality estimates in urban areas. Incorrectly interpreting home locations can lead to erroneous measurements of mobility metrics, such as the radius of gyration or average trip jump length. Section \ref{ssec:mobility_and_housing} in the Supplementary Materials underscores a few instances in mobility analyses that such errors may occur, emphasising the pertinence of incorporating housing insecurity measures into human mobility research.  The spatial relationship between housing and employment landscapes, as well as how public transit intersects with commuting behaviour, is further explored in Section \ref{sec:housing_employment_dilemma}. However, we first define the efficiency of cities' transport infrastructure to understand how urban services connect different parts of a city.

\section{Interpreting Public Transit Infrastructure \label{sec:transit}}
In the previous section, we introduced a clustering framework for identifying census tracts that are vulnerable to housing insecurity. In doing so, we are able to interpret the state of housing in various North American cities. To begin exploring whether indirect policies, such as transportation accessibility, pose further obstacles to individuals in precarious housing situations, we draw on GTFS feeds to characterise public transportation systems. We begin by defining a metric for public transportation efficiency, by comparing transit and driving times of various journeys. With this metric, we proceed to define cities based on their transit characteristics, categorising their public infrastructure as highly efficient, adequately efficient or inefficient public transportation infrastructure. Furthermore, we highlight how transit inequalities may be overlooked by neglecting to account for the spatial organisation of the housing landscape. This section aims to elucidate how transit systems serve housing demographics by measuring efficiency between and within cities and exploring residential attributes associated with proximity to transport infrastructure.

\subsection{Comparing Transit Efficiency across the USA\label{ssec:impedance}}
We begin our analysis by examining the current landscape of public transportation across all 20 cities, with a particular emphasis on how effectively transit serves each city. To investigate differences in efficiency of transportation systems, we build a transit pedestrian network using UrbanAccess and the GTFS feeds outlined in Section \ref{ssec:transit_data}. Each city's network consists of transit nodes and pedestrian nodes. Edges linking transit nodes reflect transit lines, while edges between pedestrian nodes represent paths in the road network, which is informed by OpenStreetMap. Building off these two networks, UrbanAccess connects the transit and pedestrian networks by mapping each transit node to the closest pedestrian node. Accordingly, the travel time via public transit from any two points in an urban area can be calculated as a series of transit and/or pedestrian paths. It should be noted that UrbanAccess assumes a walking speed of 3 miles per hour to calculate pedestrian travel times. 
With a given city's transit network, we can calculate the time it would take to travel from one census tract to another for a given day, during a given time frame. We construct a network for each of the 20 North American cities from 06:30AM to 10:30AM on Mondays, as that window of time captures the bulk of commutes during rush hour \cite{burd2021travel}. Transit time, alone, is not particularly informative when comparing cities of different sizes, as travel time is a function of distance and road networks. Thus, we define the efficiency of a city's transportation system by measuring how much longer a trip takes using public transit, compared to driving. We refer to this concept as {\it travel impedance}. The impedance, $imp$, from a location $x$ to location $y$ can be formally defined as:
\begin{equation}\label{eq:impedance}
\text{imp}_{x, y} = \frac{\text{transit\ time}_{x, y}}{\text{driving\ time}_{x, y}}
\end{equation}

Openrouteservice \cite{openrouteservice} provides the data for estimating the time it takes to travel from one census tract to another. A travel impedance of one implies that driving between two points takes as long as using public transit during the specified day and time range. A travel impedance, $t$, greater than one suggests that transit trips take longer than driving trips by a factor of $t$. 
To compare the efficiency of public transportation systems across cities, we define the efficiency of a city's transport system as the mean efficiency for all potential commutes (all possible pairs of census tract origins and destinations in a city). Mathematically, this is calculated by averaging the travel impedance between each pair of census tracts, where $T$ reflects the set of census tracts in a city, $c$:
\begin{equation}
\text{eff}_{c} = \frac{\sum\limits_{t_1, t_2}^{t_1,t_2\in T} imp_{t_1,t_2}}{|T|^2}
\label{eq:efficiency}
\end{equation}

\begin{figure}[H]
  \centering
  \includegraphics[width=1 \textwidth]{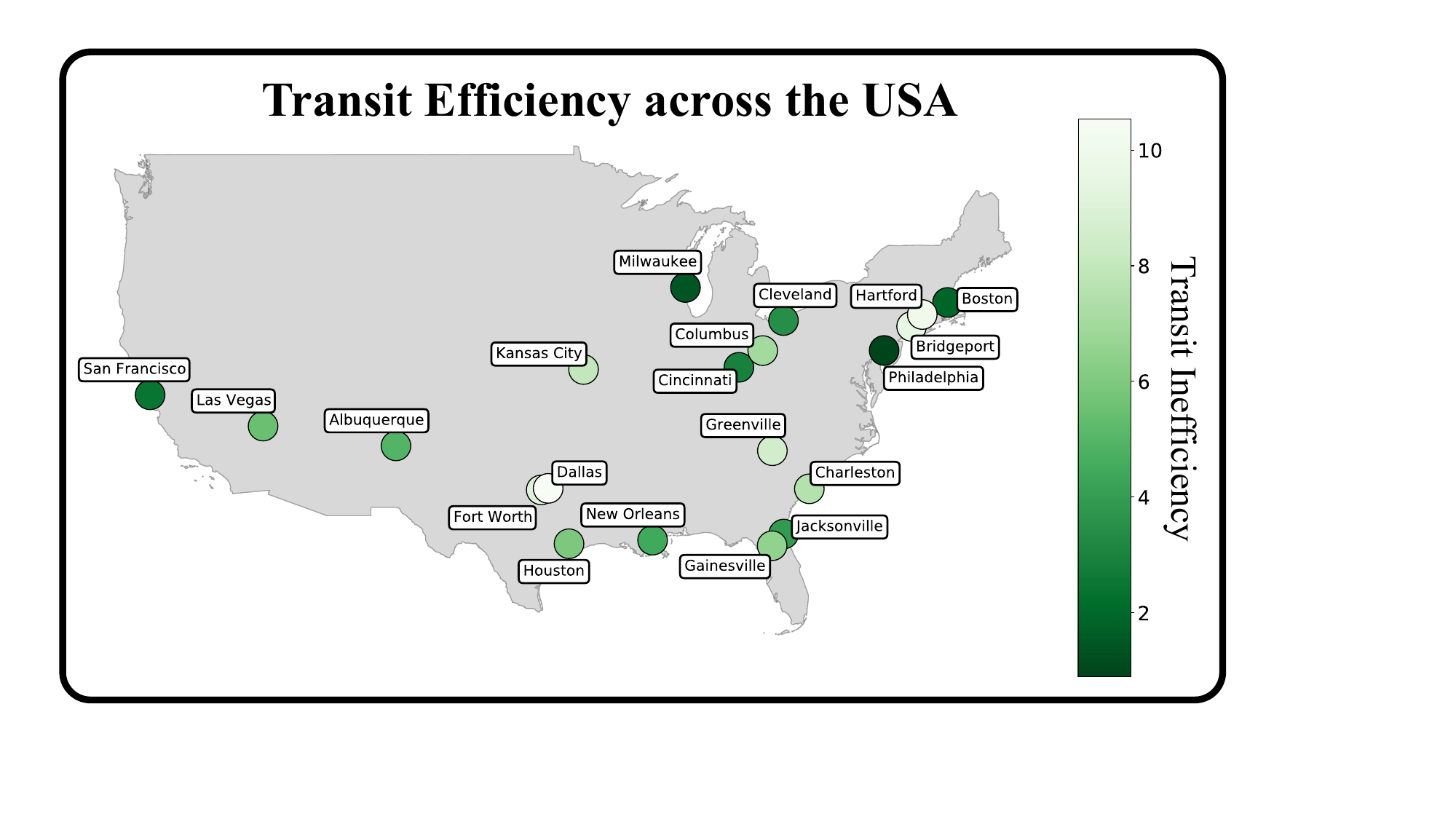}
  \caption{Map of cities in our analysis and their corresponding levels of transit efficiency. Darker hues of green indicate more efficient public transport systems.}
  \label{fig:transit_efficiency_map}
\end{figure}

Figure \ref{fig:transit_efficiency_map} captures the effectiveness of transport infrastructure for each of the cities we analyse, calculated using Equation \ref{eq:efficiency}. Darker hues of green reflect more efficient systems, while cities with whiter hues reflect regions where using transit takes significantly longer than driving. The efficiency values range from 0.90 (Philadelphia) to 10.54 (Dallas), with a median of 3.83 and a mean of 4.002 across all cities. For further details, Table \ref{si:transit_efficiency} in the Supplementary Materials lists the corresponding transit efficiency for each city. Notably, San Francisco, Milwaukee, Philadelphia, and Boston have transit systems that, on average, serve its residents in less than double the time it takes to drive. On the other hand, the transportation in Fort Worth, Bridgeport, Hartford, and Dallas generally takes more than five times as long as driving.

\begin{figure}[H]
  \centering
  \includegraphics[width=\textwidth]{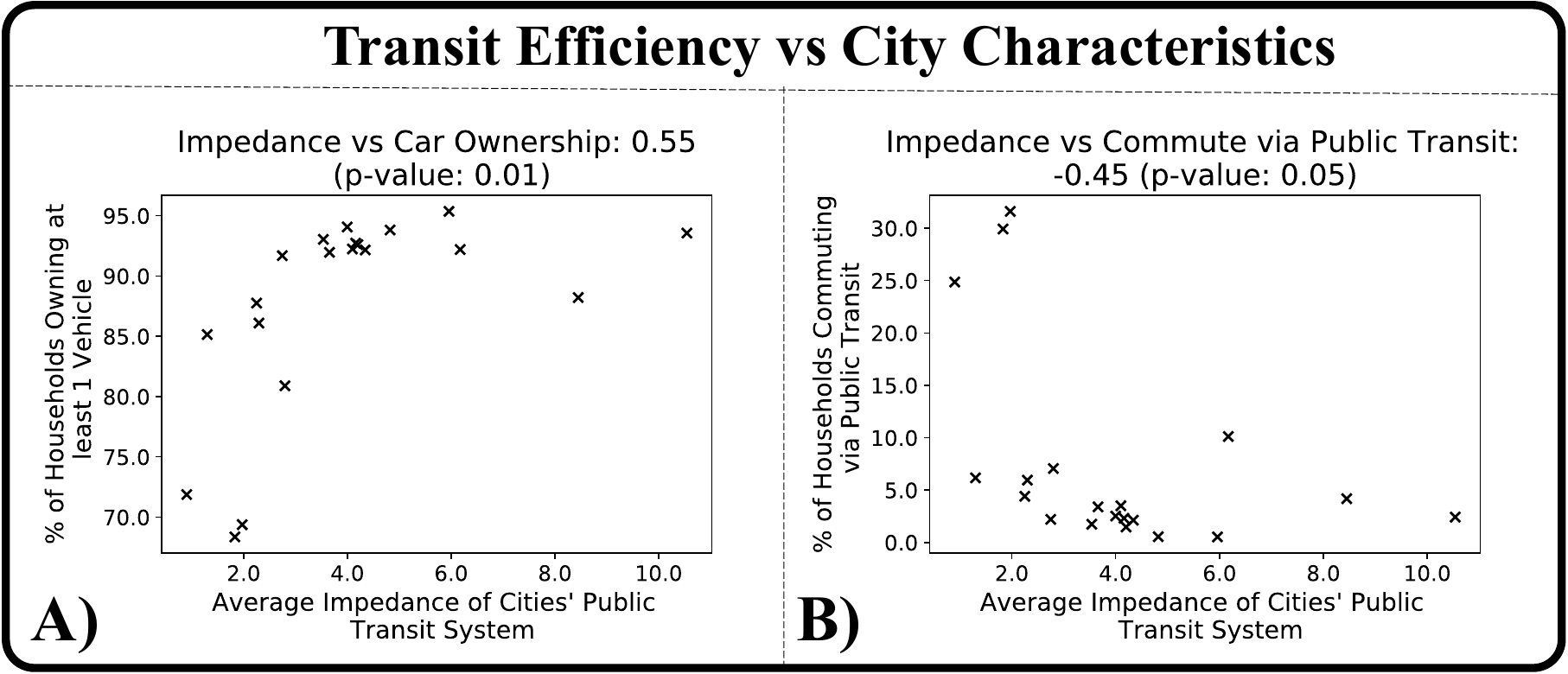}
  \caption{Exploring the relationship between transit efficiency and mobility characteristics for 20 US cities. Panel A shows a positive relationship between inefficient transit systems and the percentage of households with one or more vehicles. Panel B highlights how more efficient transit systems tend to have a higher proportion of households that commute using public transport. The Pearson correlation coefficients for car dependence and transit commutes are 0.55 and -0.45, respectively.}
  \label{fig:cityImp_carOwnership}
  
\end{figure}

With the exception of Philadelphia, the above results underscore how using transit systems in North American cities typically result in longer travel times than driving would. Moreover, Panel A in Figure \ref{fig:cityImp_carOwnership} illustrates how cities with less efficient transit systems tend to have higher rates of car ownership, with a Pearson correlation coefficient of 0.55 between the two variables. These results are consistent with research that reveals how the quality and reliability of transport infrastructure impacts the frequency with which residents use public transit \cite{wang2020impact,bordagaray2014modelling}. 

Moreover, Panel B in Figure \ref{fig:cityImp_carOwnership} shows the negative correlation (Pearson correlation coefficient of -0.45) between transit inefficiency and the percentage of the population that uses transit for commuting. Thus, we observe that less efficient transit systems are associated with a higher dependence on cars and lower levels of transit commutes. In this manner, we can reflect on how less efficient transit systems contribute to the burden of insecure housing, as the financial cost of cars depletes resources that could be otherwise invested in savings or spent on higher quality housing and choosing to commute using inefficient transit is costly from a time perspective. 

\subsection{Identifying Transit Systems that Facilitate Urban Mobility\label{ssec:transit_urban_mobility}}

\begin{figure}[H]
  \centering
  \includegraphics[width=1 \textwidth]{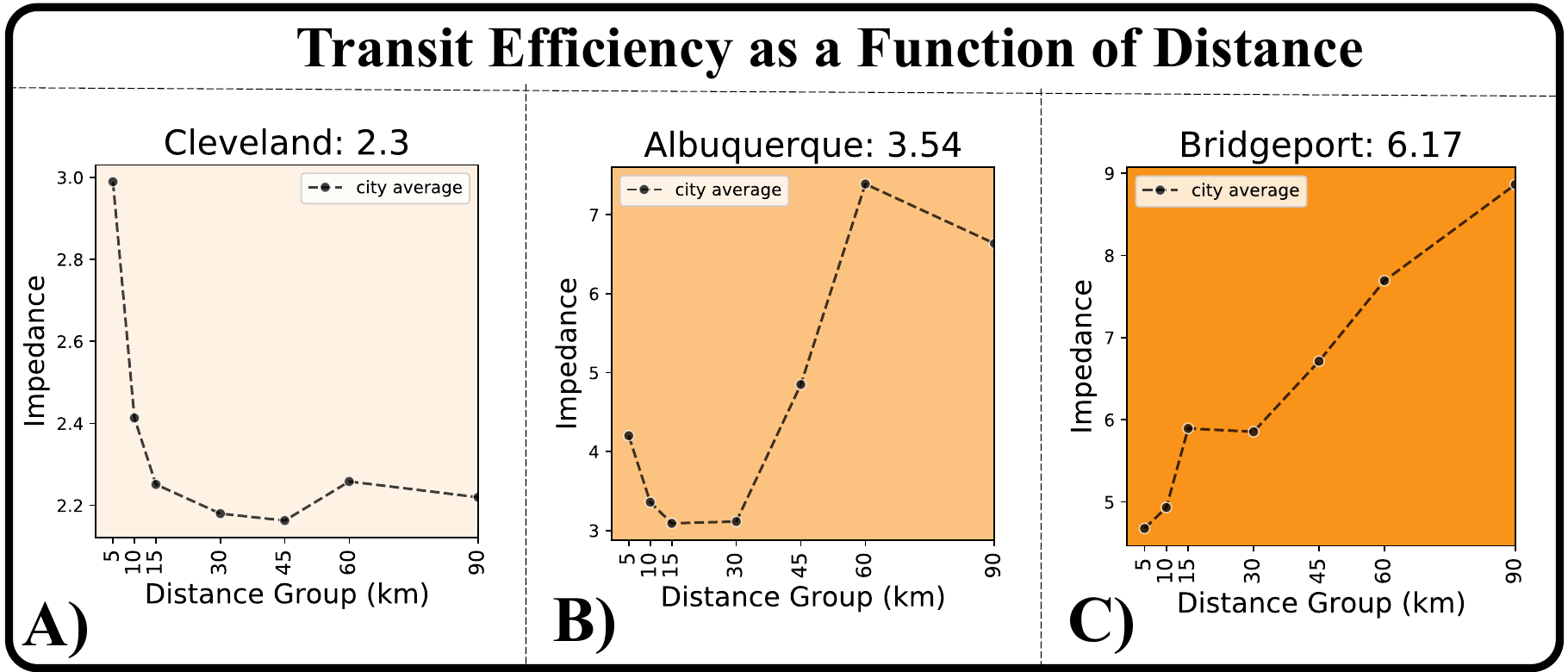}
  \caption{Examples of each of the three signatures we identify when analysing transit efficiency as a function of trip distance. Cleveland (Panel A) reflects efficient systems in which travel impedance decreases as trip distance increases. Albuquerque (Panel B) is a case of moderately efficient transit service, in which the relationship between travel impedance and trip distance switches from negative to positive at a given trip distance threshold. Panel C uses Bridgeport as a instance of inefficient transit systems, with increasing travel impedance as trip distance increases.}
  \label{fig:transitImp_tripDist}
  
\end{figure}

Our analysis of the state of public transportation in the USA has been at a city-level, allowing us to compare cities to one another. However, the cities in this analysis vary largely in size, ranging from 568 $km^2$ to 32,001 $km^2$. It should be acknowledged that there is a possibility that cities which err on the side of transit inefficiency may have effective transportation, but are larger, therefore obscuring the density and quality of the transit system. Table \ref{si:transit_efficiency} in the Supplementary Materials contains large cities that are both efficient and inefficient, indicating that region size may not be a confounding factor. To further address this potential issue, we analyse travel impedance as a function of distance. We accomplish this by creating 6 classes of transit journeys, with each category defined by how long a journey is. We refer to each class as a \textit{distance group}. 
We map each pair of census tracts to its respective distance group, based on how far the tracts are from one another. Then, for each distance group, we find the average impedance for all trips within that group. In doing so, we identify three signatures of transit efficiency with respect to trip distance, which correspond with the overall transit quality in cities. This is highlighted in Figure \ref{fig:transitImp_tripDist}, which uses Cleveland, Albuquerque, and Bridgeport to exemplify each of the discovered trends for the most, moderately, and least efficient transport systems respectively.

The first signature we observe is for the cities with efficient transit systems (lightest orange background): Philadelphia, Boston, San Francisco, Cleveland, and Jacksonville. In these cities, the travel impedance of longer journeys (30 km or more) is, on average, lower than shorter trips, indicating that the transit system is generally more efficient for trips of larger distances. These cities all tend to be more efficient, with the mean efficiency of all cities following this signature being 1.95 and the median being 1.97. The average transit impedances of these cities never exceed 3, conveying that transit times in these cities are typically upper-bounded at 3 times as long as driving times.
Another signature we unveil is for cities with inefficient public transit (darkest orange background): Gainesville, Columbus, Charleston, Kansas City, Greenville, Fort Worth, Bridgeport, Hartford, and Dallas. These cities exhibit an increasing travel impedance as the distance of trips increases, implying public transit becomes less effective than driving when journey distances increase. 

The final signature we identify is a combination of the first two signatures and is found in cities that have moderately inefficient public transportation (medium orange background): Milwaukee, Cincinnati, New Orleans, Albuquerque, Las Vegas, and Houston. These cities reveal characteristics of the first signature until a particular distance threshold. That is, travel impedance decreases as trip distances increase for shorter length trips in that region. Trips that are longer than the distance threshold follow the behaviour of the second signature, displaying increasing travel impedance with trip distance. For a more comprehensive look at the results for all 20 cities, we refer readers to Figure \ref{fig:allCities_transitImpVStripDist} of the Supplementary Materials. 

We introduce the metrics of travel impedance and transit efficiency to compare cities to one another, based on the capability of their transportation systems. We find that cities with less efficient transport infrastructure also have higher rates of car ownership. Finally, we identify different trends for how efficient trips of varying lengths are for most efficient, moderately efficient and inefficient transit systems. While this analysis highlights transit inequalities between cities, inequalities may also exist within each city, in terms of how different demographics are served by the transit system. The next section focuses on this concept, evaluating the accessibility of transit for different housing clusters based on transit efficiency and proximity.

\section{The Role of Urban Infrastructure in Facilitating Social Mobility \label{sec:housing_employment_dilemma}}
This section aims to shine light on how urban infrastructure may impose additional constraints on individuals who are already encumbered by housing insecurity. In doing so, 
we combine the housing demographics defined in Section \ref{sec:housing}, the transit characteristics outlined in Section \ref{sec:transit}, and employment data defined in Section \ref{ssec:employment_data} to illustrate how the employment landscape, housing market, and transit system hinder social mobility. To accomplish this, we investigate how public transit contributes to the burden that individuals facing housing insecurity experience, presenting additional barriers to accessing better employment opportunities.

We begin analysing the intersection of the housing and employment landscape by assessing whether areas with similar workforce characteristics express a notion of spatial proximity. Then, we identify particular census tracts that employ an unusually high concentration of its workforce from a particular housing demographic. Specifically, we use spatial autocorrelation on a global and local level to define census tracts based on the labour force that works there. This builds off of the previous sections, as we define census tracts using the residential characteristics of their employment composition to understand how employment and housing landscapes intersect. To highlight the integral role transport infrastructure plays in job accessibility, we explore how commuting times change when individuals in vulnerable housing areas start working in employment areas that provide opportunities for social mobility.

\subsection{Defining Housing and Employment Landscapes} \label{ssec:autocrrelation}

Many works have highlighted the various reasons why proximity plays a role in connecting employment and housing landscapes. Some explanations include the cost of commuting \cite{gordon2000industrial} and residential markets supplying housing to particular demographics, resulting in dense commuting flows between specific neighbourhoods \cite{hellerstein2011neighbors}. Moreover, the interconnected nature of residential and workplace segregation emphasises how housing markets and employment opportunities further contribute to experienced inequalities \cite{ellis2004work, stromgren2014factors}. Thus, we argue that of the urban and economic forces that contribute to segregated experiences, transport infrastructure should, at the very least, not add to such constraints, ideally providing sustainable alternatives to accessing better opportunities.

We apply exploratory spatial data analysis techniques on a global and a local scale. The Moran's I statistic, a common method for assessing global spatial autocorrelation, tests whether spatial clustering of a specified metric exists in a geographic data set. The metric we consider is the the workforce composition of a census tract, defined by the percentage of a workforce that is made up by a particular housing demographic. The extent of clustering is highly dependent on a spatial weights matrix, which characterises the proximity between two areas in a city. Moreover, each census tract can be defined by how much the tract's employment of a housing group deviates from the mean value, across all tracts. Thus, by combining the tract-level data with the spatial weights matrix, one can derive the degree of spatial clustering in a city, for the workforce composition of a housing demographic. This provides some insight as to whether clustering is a spatial pattern for the entire city. However, it does not define where the high rates of employment occur in the city.

To identify these clusters, we use Local Indicators of Spatial Association (LISAs) to analyse spatial autocorrelation on a local level. Thus, we can determine the census tracts that have high values of employment for a housing vulnerability group, that are also surrounded by tracts with similarly high employment rates for that demographic. In this manner, LISAs can pinpoint, what we refer to as, \textit{employment hotspots}, which indicate regions that employ a high percentage of individuals that live in a particular housing demographic. Both the local and global analysis are inferential statistics, comparing the empirical data to their randomized counterparts, in which the empirical values are maintained, but are assigned to random locations to determine the significance of spatial clustering in the data. 

Global spatial autocorrelation, in this context, assesses whether a housing group relies on particular areas of a city for job opportunities. To accomplish this, we define census tracts by the percentage of individuals working there that belong to a particular housing group. Then, we compare the employment rates of each census tract to its neighbours, characterising neighbours using Queen contiguity, in which neighbouring tracts are those that share a vertex with the focal tract. Table \ref{tab:gsa} lists the Moran's I statistic, with respect to employment rates for  each housing demographic, in which bold cells reflecting statistically significant values. For example, the second column conveys the extent of spatial concentration, with regards to how many individuals from less vulnerable residential areas makeup the workforce composition. Meanwhile, the last column captures the spatial autocorrelation of areas that employ similar rates of individuals from the most vulnerable census tracts. Table \ref{tab:gsa} is sorted by Moran's I value for the most vulnerable demographic, showing the notable role that space plays when considering the worker composition of individuals highly vulnerable to housing insecurity.

Higher values of Moran's I in the last column of Table \ref{tab:gsa} indicate that areas in a city tend to have similar employment rates of individuals that live in the most vulnerable tracts. To distinguish between areas that have high and low employment rates of each housing group, we apply local spatial autocorrelation using Local Indicators of Spatial Associations (LISAs) \cite{anselin1995local}. In this context, LISAs use the variance of employment rates and the associated spatial weights of a region to identify clusters with a high concentration of employment for a specific housing group, which are deemed employment hotspots. Panel A in Figure \ref{fig:transit_driving_times_stochastic} illustrates the housing landscape in Philadelphia and Bridgeport, with darker hues of red corresponding to census tracts that are more vulnerable to housing insecurity. Meanwhile, the purple geovisualisations convey employment hotspots for each of the housing groups, with darker hues of purple reflecting employment hotspots for individuals from the most vulnerable tracts. When we focus on the employment hotspots and residential tracts for the most vulnerable housing group, indicated by the dark purple and red, respectively, we can observe how home and workplace locations are often dependent on one another.


\subsection{Simulating Potential Avenues for Social Mobility}
\begin{figure}[H]
  \centering
  \includegraphics[width=\textwidth]{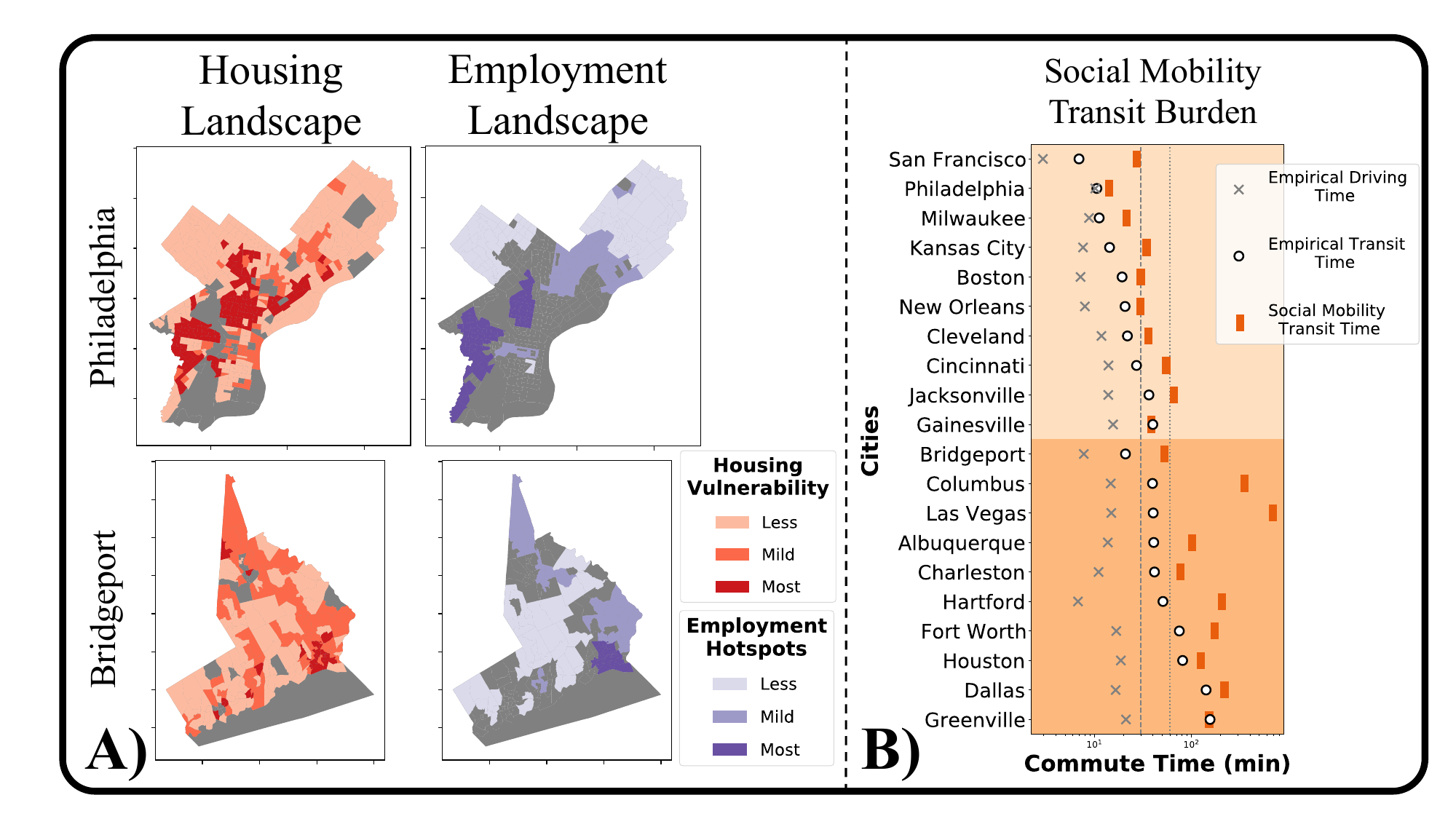}
  \caption{Panel A reflects the spatial distribution of the housing and employment landscapes in Philadelphia and Bridgeport. Darker shades of red convey higher levels of housing insecurity, while darker shades of purple reflect employment hotspots for the more vulnerable housing demographics. Panel B depicts how transit commute times change for individuals living in the most vulnerable neighbourhoods if they were start working in employment areas that would facilitate social mobility. The grey cross and white circle indicate empirical median commute times via driving and transit, respectively. Meanwhile, the orange rectangle refers to median transit commute times, averaged over 1000 of the social mobility simulations.}
  \label{fig:transit_driving_times_stochastic}
  
\end{figure}

We leverage the housing demographics, transit networks, and employment hotspots, defined in earlier sections, to examine how transit infrastructure interfaces with upwards social mobility. We focus on social mobility because wealth is often the underlying constraint preventing individuals from improving their housing conditions \cite{aurand2019gap,wardrip2011role,ramakrishnan2021housing}. We define upwards social mobility of individuals living in the most vulnerable neighbourhoods by reassigning their workplace to randomly sampled employment hotspots for the mildly vulnerable demographic. In doing so, we account for the home-workplace dependence common in urban dynamics and we assume that the hotspots for less vulnerable housing demographics provide better economic compensation. This assumption stems from the positive relationship between median household income and lower levels of vulnerability in Section \ref{ssec:housing_demographics} and shown in Panel B of Figure \ref{fig:ses_housing}. Panel B of Figure \ref{fig:transit_driving_times_stochastic} visualises how changing the workplaces of individuals commuting from the most vulnerable housing tracts impacts commuting characteristics. 

Panel B compares the commuting times for different scenarios, across each of the 20 US cities in our analysis. The grey crosses reflect median empirical driving times of individuals that both live in census tracts that are the most vulnerable to housing insecurity and commute to the most vulnerable employment hotspots. Similarly, the white circles represent the median empirical transit times for the same set of individuals. Meanwhile, the orange rectangles symbolise the median transit times for the modeled social mobility. Over 1,000 iterations, we reassign the workplaces of the same set of individuals to randomly sampled mildly vulnerable employment hotspots. We note that the x-axis is in logarithmic scale, emphasising differences between shorter commutes. The dashed line indicates a 30 minute commute, whereas the dotted line marks an hour-long commute. Moreover, we reiterate that driving times do not account for traffic, but are a reflection of the cities' road networks. Panel B underscores the dependence between housing and employment locations, as mean driving times for all cities is approximately a half hour or less. 

We observe how commuting times via transit increase for all cities, when compared to commute times using cars. However, for San Francisco, Philadelphia, and Milwaukee, transit commutes increase by less than 5 minutes. Moreover, shifting from the empirical data to the social mobility scenario reveals how commuting using public transit to areas with better opportunities leads to longer commute times, barring Greenville and Gainesville. We note that transit commutes in the social mobility simulations for Philadelphia only increases travel time by approximately 3 minutes, in comparison to its empirical counterparts. While empirical transit commuting times remain under an hour for the 16 of the 20 cities, only 10 cities maintain this characteristic in the social mobility context. Similarly, only five cities (San Francisco, Philadelphia, Boston, New Orleans and Milwaukee) have transit infrastructure that provides access to improved employment opportunities under a half-hour commute. By using housing demographics and commuting behaviour to simulate potential for social mobility, we show how the majority of cities in our analysis do not have the adequate transport service for supporting commutes, which fall under an hour-long journey, to workplaces that provide better employment opportunities.

\section{Discussion}\label{sec:discussion}

This work underscores how urban infrastructure can contribute to housing insecurity by perpetuating inequalities in how accessible areas which facilitate social mobility are. We, first, introduce a classification framework that adopts a comprehensive approach to estimating levels of housing insecurity, accounting for the various dimensions of housing conditions. Then, we use public transit and street network data to characterise US cities based on their transport infrastructure. Finally, we highlight how transit systems in most cities pose obstacles to accessing employment areas that are associated with improved housing conditions. In this manner, we show how urban infrastructure impedes individuals' abilities to live in improved housing for various cities.

In Section \ref{sec:housing} we combine census and eviction data to estimate levels of housing insecurity on a census tract level in 20 US cities. Considering Cox's definition of housing insecurity, we quantify housing insecurity with respect to affordability, quality, and stability \cite{cox2019road}. Existing approaches to defining housing insecurity, in the context of urban analytics, include using a specific housing feature as a rough proxy, such as rent burden or forced moves \cite{hepburn2023beyond,boeing2020online}. Others have multiple housing features characterising one dimension,  We focus on Cox's definition as it captures financial, structural, and social forces that influence the state of housing. It is important to note that the seven dimensions proposed by Cox stem from a Global North perspective, with its definition based on US housing policies. Attempts to develop a comprehensive measure for Global South incorporate features such as sanitation and water access \cite{roy2020exploratory}. The distinction between these two definitions is imperative, considering that different histories, cultures, and environments can re-frame the relevance of a housing dimension, and the features that can be used to estimate said dimension. Thus, the accompanying code to generate housing clusters must be used with great care and in the proper context. Moreover, this work is limited in data availability of housing conditions. Thus, the neighbourhood and homelessness dimensions are yet to be incorporated. However, given the flexibility of the proposed framework, introducing these dimensions is simply a matter of modifying the rank-based approach to. Potential neighbourhood characteristics can be defined using crime data sources for safety or built form metrics for quality. Ultimately, our approach aims to capture various mechanisms that contribute to poor housing experiences, which we validate by comparing to a range of socioeconomic characteristics such as income, educational attainment, and mobility behaviour.

Furthermore, our work leverages open source tools to define travel impedance based on how much longer a journey takes using transit than by car. By averaging travel impedance over potential trips in a city, we identify Philadelphia, Milwaukee, and San Francisco as the three US cities with the most efficient transit systems, of those considered. Moreover, we observe three types of transit systems based on how transit efficiency relates to trip distance. In line with research that demonstrates the decreasing significance of distance due to improved transit systems, we find that the cities with the most efficient transit service overall tend to have equally, if not more, efficient transit impedance for trips of longer distances \cite{pirie2009distance}.

Finally, by incorporating mobility behaviour between residential and work areas, we unveil how transit infrastructure can impose additional hurdles to accessing workplaces that provide better financial opportunities. While studies have shown that targeted efforts in improving transit access to job opportunities has a positive effect on individual employment probability and individual income, particularly improving employment probabilities for lower income individuals \cite{saif2019public, bastiaanssen2022does}. However, we explore the geospatial layout of employment opportunities with residential landscapes to see how these efforts may also perpetuate inequalities in accessing jobs with different characteristics. Thus, this analysis contributes to research that motivates exploring inequality analyses from a spatial perspective, emphasising the importance of space and the built environment in social processes. Ultimately, housing conditions impact the level of comfort and belonging individuals experience within their environment \cite{easthope2004place, ozer2022changing}. Thus, we aim to highlight how the strain of housing insecurity is exacerbated by urban features that can hinder vulnerable populations from breaking out of the cycle of poverty.

\bibliography{sn-article}

\end{document}


\maketitle
\section{Background \label{sec:background}}

\subsection{Human Mobility and Housing Insecurity \label{ssec:mobility_and_housing}}
Take, for example, the radius of gyration, which, in the context of human mobility, measures the linear distance that characterises individuals' trajectories in a given time period \cite{liu2018temporal, gonzalez2008understanding, pappalardo2013understanding}. This is measured using mobility trajectories to compare the distance between a user's location and the centre of mass of their trajectory. By assuming static residency for individuals that move houses during the span of the data period, the centre of mass would conflate mobility characteristics before and after the forced move. This notion can be extended to many other metrics in human mobility. It is common practice to estimate residential location by the most frequent location that an individual is in during the very early hours of the morning \cite{ahas2010using, xu2015understanding, xu2018human}. This approach assumes static residency as well. In the case of a forced move, a user may, at first, frequently appear in one location during the morning, but then shift to another home location after the forced move. Using the standard home estimation techniques, the home location would be set to whichever house the individual was living at for longer during the time period of the data. Then, any identified mobility differences across demographics, such as average travel time or trip length, could be inaccurate, especially when considering demographics that are particularly vulnerable to housing instability. Thus, it becomes clear that inequality studies in human mobility must extend beyond sociodemographic traits of a population and incorporate urban stressors such as housing insecurity.

\subsection{Measuring Housing Insecurity \label{ssec:housing_insecurity}}
Social scientists and urban planners have consistently and carefully studied the sources and consequences of unstable housing using different granularities. In particular, the COVID-19 pandemic enforced the importance of studying housing insecurity. As travel restrictions limited mobility, individuals were confined to their places of residence, emphasising the significance of housing quality \cite{peters2020our, benfer2020covid}. Over the years, understanding the motivation for residential movements of disadvantaged groups has posed a considerable challenge. Desmond et. al suggest eviction, coupled with neighbourhood dissatisfaction, gentrification, and slum clearance as potential explanations for the high levels of moves among the urban poor \cite{desmond2012eviction}. Housing insecurity has been shown to have a negative effect on job accessibility, wellbeing, and the stability of support networks \cite{desmond2016housing, desmond2015eviction, kingsley2009impacts}.

Ruonavaara points to the 'linguistic ambiguity' of housing as an indication of its complexity \cite{ruonavaara2018theory}. The dualistic nature of 'housing' allows it to shape-shift between verb and noun forms. Following this trajectory, developing a definition for [adequate] housing is not only bottlenecked from a linguistic perspective, but a political, economic, and, of the greatest relevance to this research, a measurable one. Developing a consistent metric for housing insecurity has been a long-standing obstacle \cite{leopold2016improving}. In an effort to provide a universal metric for housing insecurity, Cox et al. survey existing research in housing issues to define seven dimensions of housing insecurity: housing stability, housing affordability, housing quality, housing safety, neighbourhood safety, neighbourhood quality, and the last, optional dimension, homelessness \cite{cox2019road}. Housing stability focuses on concepts such as overcrowding in houses, evictions, and frequent moves. Housing affordability encompasses financial aspects of housing such as rent burden, incomplete or late payments, mortgage, and taxes. Meanwhile, housing quality focuses on the robustness of the house as a physical structure. This includes characteristics such as the functionality of appliances and how rundown a house's interior and exterior are. Housing safety differs from housing quality, in that is measures the presence of vital housing facilities such as heating, water access, and pests. The neighbourhood level characteristics portray the notion that residential locations are not based solely on the house, but also account for the environment in which the house exists. Neighbourhood safety is described by crime rates, abandoned buildings, proximity to environmental threats, noise levels, and traffic. Neighbourhood quality encompasses urban infrastructure quality, such as amenity accessibility. We use these dimensions, which capture different facets of housing insecurity, to define the state of housing for different urban areas in the United States of America (USA).

\section{Data Sources \label{sec:si_data_souurces}}
\subsection{Eviction Data\label{ssec:eviction_data}}
In the case of Dallas, Texas, the data published by the Eviction Lab was sourced from the North Texas Eviction Project, which is a branch of the Child Poverty Action Lab. Theoretically, these two data sources should be the same, however, the data cleaning methods implemented by the Eviction Lab may account for the differences in eviction rates seen in Figure \ref{fig:evictions_dallas}. They conducted this process, which consisted of identifying and removing duplicate entries and modifying blatantly incorrect addresses, to allow for data comparison across different cities \cite{desmond2018eviction}. Although we expect eviction rates to be underestimated, they provide clearer insight into differences in eviction rates across census tracts and cities and elucidate temporal patterns in evictions.

\subsection{Housing Data Sources\label{ssec:si_housing_data}}
\subsubsection{Housing Affordability \label{sssec:housing_affordability}}
To define the dimension of Housing Affordability, we use census tract-level data from the following ACS tables: (a) B25070: Gross Rent as a Percentage of Household Income (b) B25097: Mortgage Status by Median Value and (c) B25001: Housing Units. Rent as a Percentage of Household Income helps to define rent burdened households, which the U.S. Department of Housing and Urban Development (US HUD) define as households that spend 30\% or more of their income on rent \cite{steffen2015worst}. They extend this definition, labelling households at least 50\% of their income on rent as severely rent-burdened. It should be noted that the percentage for individuals experiencing a net loss is not computed, potentially neglecting an extreme case of economic depravity. We normalise the number of severely rent burdened households in each tract with respect to the total number of surveyed households in that tract. Moreover, we measure a census tract's housing stock by dividing the number of housing units by the population of each tract.

\begin{figure}[H]
  \centering
  \includegraphics[width=1 \textwidth]{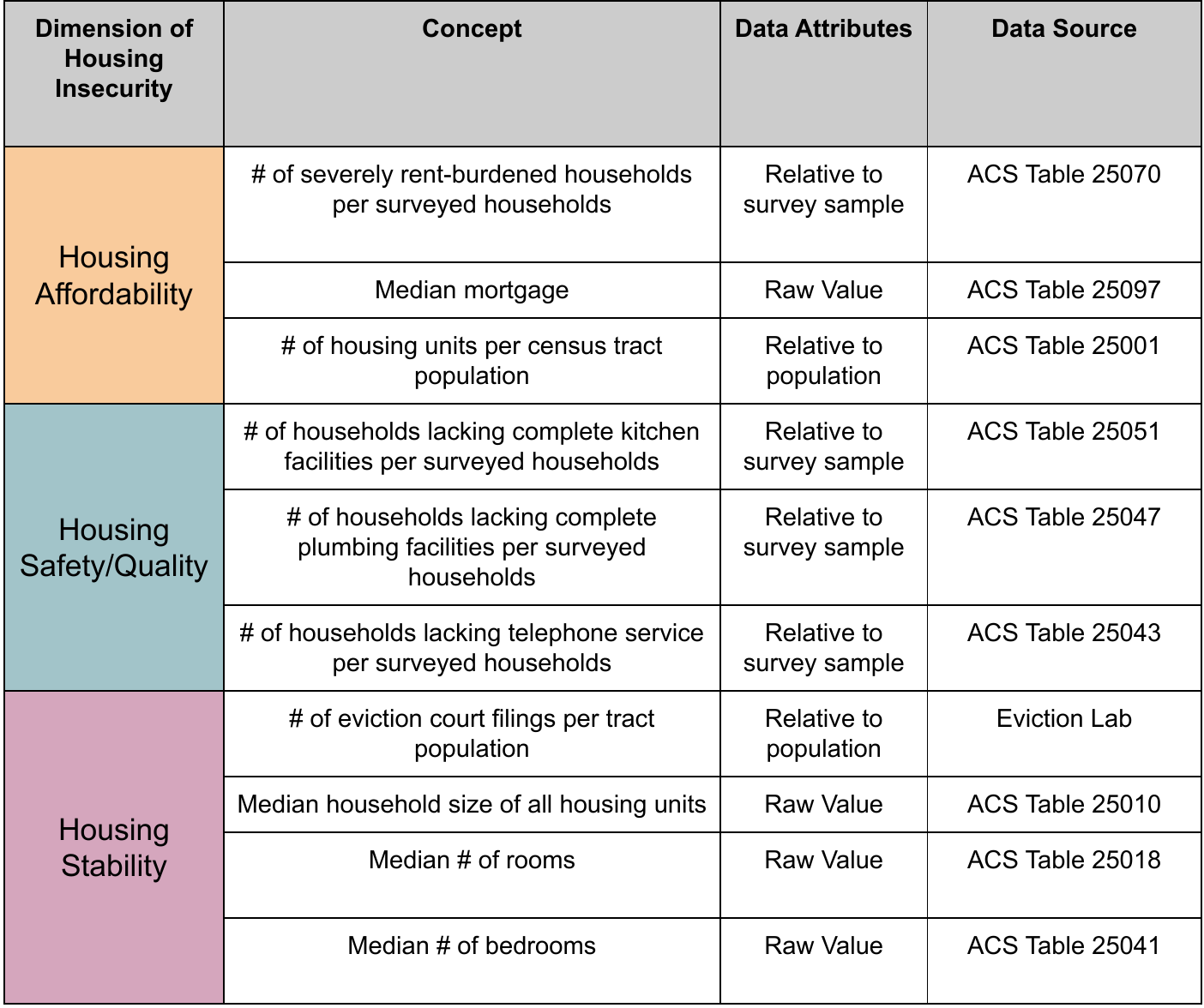}
  \caption{Summary of data sources for the three housing dimensions used in the clustering framework. Most of the data is provided by the US Census American Community Survey, with the Eviction Lab providing data on formal eviction rates to characterise the housing stability dimension. }
  \label{fig:housing_data_summary}
  
\end{figure}

\subsubsection{Housing Quality and Safety \label{sssec:housing_quality}}
Due to the range of data available from the ACS, we combine Housing Quality and Housing Safety into one dimension. Other potential data sources include the American Housing Survey and data hubs managed by individual counties or cities. The American Housing Survey provides data on neighbourhood safety and quality, but the finest-resolution it provides is metropolitan areas. On the other hand, city-managed data sources provide highly detailed accounts of housing complaints, but are limited by how organised and transparent each city is, introducing difficulties when expanding the analysis to different regions. Thus, we use the following tables from the ACS to define the quality and safety of housing: (a) B25051: Kitchen Facilities for All Housing Units, (b) B25047: Plumbing Facilities for All Housing Units, and (c) B25043: Tenure by Telephone Service Available by Age of Householder. Households must contain a sink, a stove, and a refrigerator to be considered as having complete kitchen facilities. 
Similarly, they must have hot and cold running water, a flush toilet, and a bathtub or shower to qualify as a household with complete plumbing facilities. Telephone service provides one way of measuring household isolation. This is pertinent for providing adequate medical and crime-related services to households. For a household to be considered as having service available, they must have access to telephone service and have a functional phone. All of the above data is measured, for each census tract, as a fraction of households without these services over the total number of households surveyed. 

\subsubsection{Housing Stability \label{sssec:housing_stability}}
Finally, to quantify Housing Stability, we draw on eviction and overcrowding data. The AHS does not provide eviction data, leaving us to depend on county and city-level governments to make the data accessible. Cities like New York City, Dallas, Detroit, and San Francisco have processed eviction court filings to publish data referring to evictions. This introduces a significant array of issues. First and foremost, eviction data assumes that each eviction is processed through a judicial means. However, landlords often enforce informal evictions as they provide a more affordable means for the same outcome. This includes changing the locks or paying a family to move \cite{hartman2003evictions}. Other informal evictions range from cities declaring the housing inhabitable or the threat of foreclosure \cite{been2009tenants}. Documented evictions also impact tenants' abilities to rent in high quality neighbourhoods, as landlords often check prospective tenants' housing history, giving landlords an often underestimated influence in neighbourhood composition and gentrification \cite{desmond2015forced}. Yet another issue is data processing from the courts' perspective. Errors in data entry and nuanced rulings lead to inaccurate measures of evictions when compounded into accessible tables \cite{porton2021inaccuracies}. Despite all of these issues, leveraging eviction court filings is the most promising option for estimating forced moves. Projects like Anti-Eviction Mapping Project and the Eviction Lab are vocal about increasing eviction data availability. While both projects work with local organisations, the former provides a broader range of analyses and case studies spanning many North American cities. Meanwhile, the latter focuses on how evictions were impacted by COVID-19 and publish weekly and monthly eviction counts for various cities on a census tract and zip code tabulation area level. For this study, we focus on cities for which the Eviction Lab has processed eviction rates on a census tract level. 

We return to the ACS to estimate overcrowding in housing. Drawing on studies by the US HUD, which measure overcrowding using Persons-Per-Room (PPR), Persons-Per-Bedroom (PPB), and Unit-Square-Footage-Per-Person, we use the following ACS tables: (a) B25010: Average Household Size of Occupied Housing Units by Tenure, (b) B25018: Median Number of Rooms and (c) B25041: Bedrooms \cite{blake2007measuring}. These tables capture PPR and PPB. However, the ACS does not provide data on the physical size of housing units. The American Housing Survey, which we previously alluded to in Section \ref{sssec:housing_quality}, does provide physical housing characteristics, but only on a metropolitan level, which is not fitting for the granularity of our studies. Thus, we capture Housing Stability by combining eviction filing data from the Eviction Lab and overcrowding metrics, such as PPR and PPB, derived from the ACS.

\section{Data Preparation and Clustering \label{sec:si_data_prep_and_clustering}}
\subsection{Data Preprocessing \label{ssec:data_preprocessing}}
This section outlines how we preprocess the housing data introduced in Section \ref{sec:data} of the main manuscript and elaborated on in Section \ref{sec:si_data_souurces} of this document. We begin by replacing all negative values with signaling NaNs. Typically, negative values appear in the American Community Survey to indicate insufficient data, due to sample sizes, for example. 
 Then, we retain the data from census tracts that have a population greater than zero and that have data on at least one of the housing affordability features (rent burden, mortgage, or housing stock). In doing so, we only consider census tracts in which individuals reside. We standardise the data such that, for a given city, each housing feature has a mean value of zero and a standard deviation of one. This is done to ensure that features with larger raw values (i.e. mortgage) will not have a greater influence on the results than features with smaller values (i.e. evictions per capita). 

\subsection{Details on Spectral Clustering \label{ssec:spectral_details}}
We implement Spectral Clustering to measure levels of housing insecurity within different urban areas. Among the selection of various unsupervised learning techniques, we rely on Spectral Clustering as it makes minimal assumptions about the shape of data, while approaches such as K-Means assumes the data is spherical. Moreover, other methods are extremely sensitive to initialisation so require numerous starts to derive high quality clusters. The process of Spectral Clustering can be broken down into three main steps. The first step entails constructing an $|N| \times |N|$ similarity matrix where $|N|$ is the number of records. This can be done using a radial basis function, which applies a Gaussian function the Euclidean distances between each data point. Since each node in our graph represents the housing characteristics of a particular census tract, we tend to have high values of $|N|$. Thus, we use a K-Nearest Neighbors similarity graph to create a more sparse matrix. The second step uses properties of the Graph Laplacian matrix to create a low-dimensional spectral embedding of the affinity matrix created in the first step. The diagonal matrix of degrees, $D$, is necessary to calculate the Laplacian Matrix, $L$, for an affinity matrix, $A$. $D$ is simply an identity matrix, where each diagonal value is the row-wise summation of edge weights for the respective node in the graph. The Laplacian Matrix can, then, be defined in numerous ways. For the approach of identifying different housing vulnerability clusters, we use the Graph Laplacian, which is calculated by subtracting the diagonal matrix, $D$, from the affinity matrix, $A$. In this manner, all values on the diagonal of $L$ capture the weighted degree of each node, while each cell $l_{ij}$ is the negative edge weight, if an edge exists, or zero otherwise. One of many fascinating properties of the Laplacian matrix is that its rows and columns sum to zero. Moreover, the eigenvalues of a graph's Laplacian matrix informs structural properties of the graph, namely the number of components it has. If the first $x$ eigenvalues are zero, this indicates that the respective graph built from $A$ has $x$ connected components. Eigenvalues that are near-zero indicate a loosely-connected graph, which couldn't be split into separate components with minimal cuts. Conversely, a very dense graph would have eigenvalues near $|N|$. The last step involves applying a classical clustering algorithm, commonly K-Means to the embedding to create partitions in the data. It becomes clear, then, that eigenvalues can be leveraged to identify the number of clusters into which the data, embedded into a graph, can be clustered.


\subsection{Socio-demographic Characteristics of Housing Groups\label{ssec:housing_demographics}}

\begin{figure}[H]
  \centering
  \includegraphics[width=1 \textwidth]{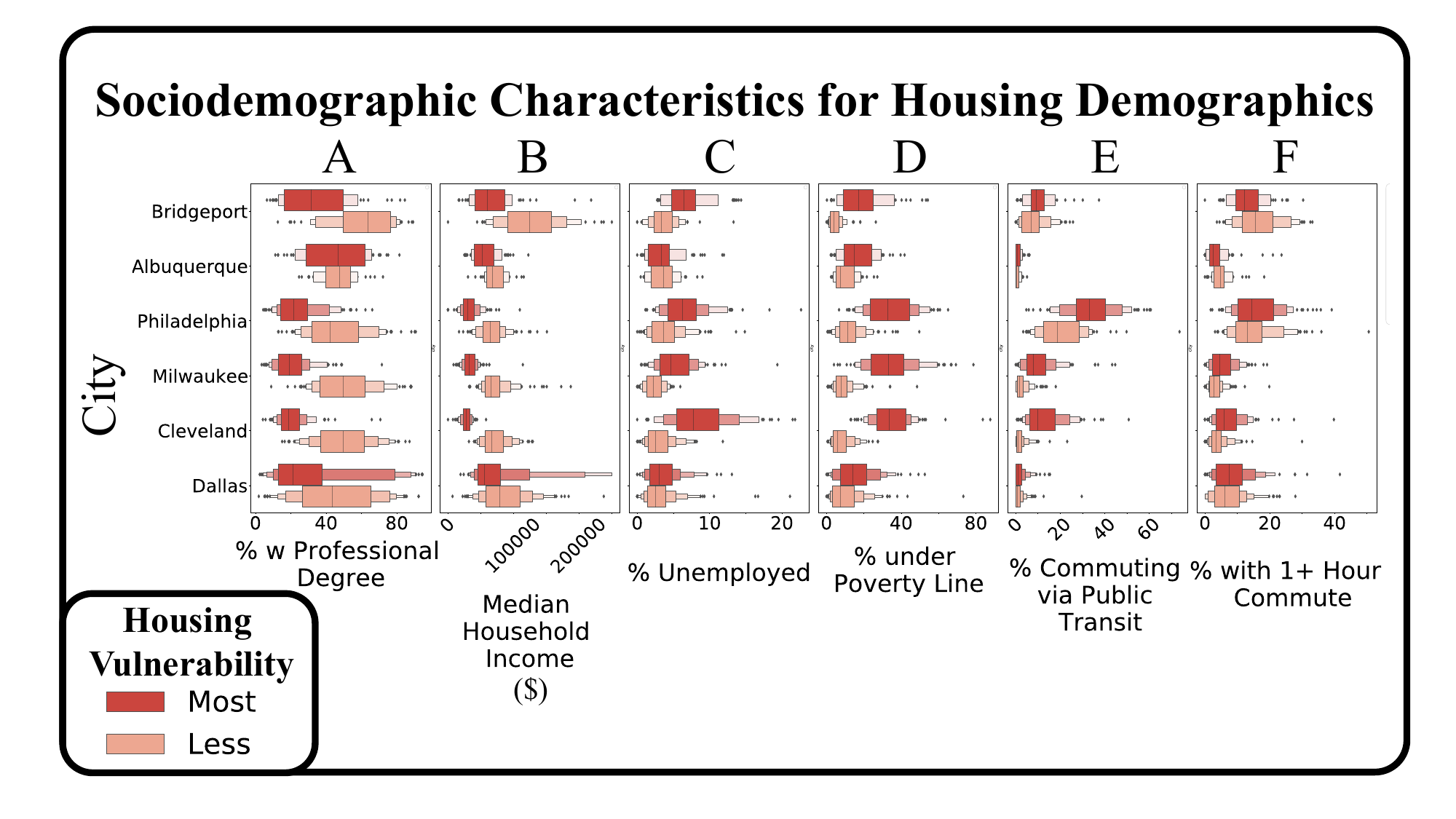}
  \caption{Comparing socioeconomic characteristics of the most and less vulnerable housing demographics for Bridgeport, Albuquerque, Philadelphia, Milwaukee, Cleveland, and Dallas. This Figure conveys how the most vulnerable neighbourhood tend to have the lower rates of educational attainment (Panel A), lower median households incomes (B), higher unemployment (Panel C), higher poverty (Panel D), more dependence on transit for commuting (Panel E) and higher percentages of commutes over an hour long (Panel F).}
  \label{fig:ses_housing}
\end{figure}

We validate our clustering approach by exploring how the housing demographics we defined in the previous section relate to socioeconomic variables associated with employment, wealth, and commuting. Specifically, with respect to each tract, we gather census data capturing (a) the percentage holding a professional degree, (b) the median household income, (c) unemployment rates, (d) poverty rates, (e) the percentage commuting using public transportation, and (f) the percentage of the tract with a commute longer than an hour. Figure \ref{fig:ses_housing} compares these socioeconomic variables across the tracts in the less vulnerable and most vulnerable housing groups for Bridgeport, Albuquerque, Philadelphia, Milwaukee, Cleveland, and Dallas. Meanwhile, Tables \ref{tab:ses_eduAttainment} to \ref{tab:ses_ptCommuteTime} list the sociodemographic characteristics of the most and less vulnerable housing demographics for all cities. We choose these 6 cities as they span a range of geographic and public transport characteristics. The tables detailing statistical properties for all cities can be found in the Supplementary Materials. 

Panel A in Figure \ref{fig:ses_housing} illustrates how census tracts that are the least vulnerable to housing insecurity tend to have a higher fraction of its residents holding a professional degree. This considers Associate's, Bachelor's, Master's, Professional school, and Doctorate degrees as qualifications for estimating the level of educational attainment in a region. Similarly, Panel B highlights how the most vulnerable neighbourhoods have lower median household incomes than the census tracts that are less vulnerable. On the other hand Panels C and D consider deprivation indicators such as unemployment and poverty. In both these cases we see associations between housing vulnerability and higher rates of unemployment and poverty, for all of the 6 cities considered. When we analyse commuting characteristics of housing demographics, we generally see more dependence on transit systems for the most vulnerable neighbourhoods. We note that Albuquerque and Dallas have much lower percentages of transit commutes than the other 4 cities. Finally, when considering commute time, irrespective of commute mode, we observe less distinct disparities between the housing demographics. We hypothesize that this has to do with spatial organisation of residential and employment areas for each housing demographic. That is individuals organise their residential and employment locations to be close to one another, an idea that is consistent in urban commuting literature \cite{hickman2010planning}.

\begin{table}[]
    \caption{Statistical properties of educational attainment levels in the census tracts that are less and most vulnerable to housing insecurity. Generally, the most vulnerable tracts tend to have a lower fraction of their population holding an Associate's Degree or higher.}
    \centering
    \begin{tabular}{l|ll|ll|ll}
        
        \toprule
        \multicolumn{7}{c}{\textbf{\% holding an Associate's Degree or higher}} \\
        \midrule

        {} & \multicolumn{2}{l}{Mean} & \multicolumn{2}{l}{Median} & \multicolumn{2}{l}{Std Dev} \\
        {Housing Vulnerability} & \textit{Less}  & \textit{Most}  & \textit{Less}  & \textit{Most}  & \textit{Less}  & \textit{Most}  \\
        City         &                                           &                 &                 &                 &                 &                 \\
        \midrule
        Albuquerque  &                                     46.90 &           45.56 &           47.75 &           46.79 &           11.40 &           18.25 \\
        Boston       &                                     59.39 &           64.36 &           57.92 &           74.99 &           19.24 &           26.35 \\
        Bridgeport   &                                     60.27 &           34.93 &           63.48 &           31.66 &           18.20 &           20.35 \\
        Charleston   &                                     60.74 &           46.44 &           62.51 &           41.34 &           11.50 &           25.42 \\
        Cincinnati   &                                     57.52 &           37.29 &           56.93 &           31.14 &           17.30 &           19.83 \\
        Cleveland    &                                     49.80 &           20.84 &           49.78 &           18.56 &           16.19 &           10.02 \\
        Columbus     &                                     40.09 &           23.42 &           38.07 &           20.33 &           17.47 &           12.70 \\
        Dallas       &                                     46.18 &           31.56 &           43.47 &           21.35 &           23.02 &           26.14 \\
        Fort Worth   &                                     53.73 &           29.02 &           53.58 &           29.92 &           15.83 &           15.82 \\
        Gainesville  &                                     60.73 &           56.21 &           65.46 &           57.08 &           15.98 &           17.48 \\
        Greenville   &                                     49.02 &           31.32 &           54.57 &           30.53 &           19.17 &           13.34 \\
        Hartford     &                                     50.82 &           22.89 &           51.73 &           20.12 &           12.43 &           14.34 \\
        Houston      &                                     54.66 &           28.89 &           55.42 &           21.21 &           20.42 &           20.78 \\
        Jacksonville &                                     42.88 &           25.73 &           43.95 &           23.95 &           14.03 &           10.89 \\
        Kansas City  &                                     37.84 &           23.56 &           36.03 &           22.57 &           13.95 &           11.36 \\
        Las Vegas    &                                     41.71 &           23.25 &           40.98 &           21.91 &           13.31 &           12.02 \\
        Milwaukee    &                                     50.89 &           20.77 &           49.81 &           19.32 &           17.17 &           10.88 \\
        New Orleans  &                                     53.65 &           29.83 &           60.36 &           28.96 &           21.45 &           14.41 \\
        Philadelphia &                                     45.40 &           24.28 &           42.23 &           21.64 &           18.09 &           13.27 \\
        St Louis     &                                     60.23 &           20.55 &           60.21 &           17.24 &           11.46 &           10.79 \\
        \bottomrule
    \end{tabular}
    \label{tab:ses_eduAttainment}
\end{table}

\begin{table}[]
    \caption{Statistical properties of median income levels in the census tracts that are less and most vulnerable to housing insecurity. Generally, the most vulnerable tracts tend to have a lower household incomes than the less vulnerable tracts.}
    
    \centering
    \begin{tabular}{l|ll|ll|ll}
        \toprule
        \multicolumn{7}{c}{\textbf{Median Household Income}} \\
        \midrule
        {} & \multicolumn{2}{l}{Mean} & \multicolumn{2}{l}{Median} & \multicolumn{2}{l}{Std Dev} \\
        {Housing Vulnerability} & \textit{Less}  & \textit{Most}  & \textit{Less}  & \textit{Most}  & \textit{Less}  & \textit{Most}  \\
        
        City         &                         &                 &                 &                 &                 &                 \\
        \midrule
        Albuquerque  &                   71708 &           55928 &           67895 &           51890 &           19206 &           21165 \\
        Boston       &                   97527 &           77162 &          100058 &           71994 &           40271 &           43245 \\
        Bridgeport   &                  126905 &           67659 &          124858 &           60455 &           48872 &           38435 \\
        Charleston   &                   85666 &           66016 &           83493 &           41307 &           33447 &           48145 \\
        Cincinnati   &                   85413 &           41643 &           79862 &           38001 &           36104 &           19433 \\
        Cleveland    &                   71616 &           27990 &           66994 &           28093 &           20978 &            8300 \\
        Columbus     &                   67694 &           37739 &           68750 &           36342 &           18651 &           13648 \\
        Dallas       &                   85031 &           76001 &           78842 &           55765 &           36834 &           53783 \\
        Fort Worth   &                  101066 &           59572 &           92727 &           55365 &           41749 &           24393 \\
        Gainesville  &                   67019 &           39693 &           68308 &           36271 &           31781 &           16974 \\
        Greenville   &                   80690 &           48400 &           73170 &           48823 &           23629 &           13471 \\
        Hartford     &                   83640 &           37472 &           82554 &           34934 &           18296 &           13455 \\
        Houston      &                  105268 &           58296 &           98194 &           51619 &           45645 &           29536 \\
        Jacksonville &                   70382 &           39627 &           68286 &           36886 &           19119 &           13891 \\
        Kansas City  &                   69273 &           38012 &           66070 &           37462 &           21925 &           10902 \\
        Las Vegas    &                   84223 &           47550 &           76366 &           43400 &           25637 &           19456 \\
        Milwaukee    &                   71907 &           33607 &           65950 &           32487 &           25398 &           12989 \\
        New Orleans  &                   64826 &           31005 &           63778 &           28981 &           36642 &           14952 \\
        Philadelphia &                   66942 &           32504 &           64966 &           29697 &           22345 &           15818 \\
        St Louis     &                   61021 &           29586 &           62014 &           27038 &           10782 &            8837 \\
        \bottomrule
        \end{tabular}
    \label{tab:ses_income}
\end{table}

\begin{table}[]
    \caption{Statistical properties of unemployment levels in the census tracts that are less and most vulnerable to housing insecurity. Generally, the most vulnerable tracts tend to have a higher unemployment rates, with the two exceptions being Albuquerque and Boston.}

    \centering
    \begin{tabular}{l|ll|ll|ll}
        \toprule
        \multicolumn{7}{c}{\textbf{\% Unemployed}} \\
        \midrule
        {} & \multicolumn{2}{l}{Mean} & \multicolumn{2}{l}{Median} & \multicolumn{2}{l}{Std Dev} \\
        {Housing Vulnerability} & \textit{Less}  & \textit{Most}  & \textit{Less}  & \textit{Most}  & \textit{Less}  & \textit{Most}  \\
        City         &              &      &        &      &         &      \\
        \midrule
        Albuquerque  &         3.66 & 3.59 &   3.64 & 3.28 &    2.17 & 2.67 \\
        Boston       &         4.04 & 3.75 &   3.34 & 2.97 &    3.44 & 3.00 \\
        Bridgeport   &         3.65 & 6.87 &   3.37 & 6.40 &    2.11 & 3.17 \\
        Charleston   &         1.98 & 2.90 &   1.63 & 2.54 &    1.47 & 1.93 \\
        Cincinnati   &         2.47 & 5.75 &   2.17 & 4.84 &    1.59 & 4.00 \\
        Cleveland    &         3.06 & 8.69 &   2.50 & 7.76 &    2.06 & 4.60 \\
        Columbus     &         3.00 & 5.88 &   2.63 & 5.31 &    2.38 & 3.43 \\
        Dallas       &         3.12 & 3.52 &   2.54 & 3.03 &    2.73 & 2.43 \\
        Fort Worth   &         2.50 & 4.06 &   2.26 & 3.59 &    1.65 & 2.73 \\
        Gainesville  &         2.57 & 4.47 &   2.50 & 4.39 &    1.79 & 2.86 \\
        Greenville   &         1.81 & 3.62 &   1.63 & 3.09 &    1.22 & 1.81 \\
        Hartford     &         2.83 & 6.59 &   2.34 & 6.33 &    2.11 & 2.97 \\
        Houston      &         3.56 & 4.43 &   3.07 & 3.99 &    2.51 & 3.01 \\
        Jacksonville &         2.88 & 4.66 &   2.47 & 4.12 &    1.94 & 2.49 \\
        Kansas City  &         3.09 & 4.33 &   2.71 & 3.81 &    2.30 & 3.18 \\
        Las Vegas    &         3.85 & 5.25 &   3.50 & 4.82 &    2.23 & 2.74 \\
        Milwaukee    &         2.34 & 5.19 &   2.19 & 4.62 &    1.33 & 2.89 \\
        New Orleans  &         4.23 & 7.39 &   3.80 & 6.86 &    3.25 & 5.05 \\
        Philadelphia &         4.03 & 6.65 &   3.63 & 6.28 &    2.65 & 3.41 \\
        St Louis     &         2.27 & 6.80 &   1.85 & 6.94 &    1.71 & 3.06 \\
        \bottomrule
        \end{tabular}
    \label{tab:ses_unemployment}
\end{table}

\begin{table}[]
    \centering
    \caption{Statistical properties of poverty levels in the census tracts that are less and most vulnerable to housing insecurity. The most vulnerable tracts tend to have much higher rates of poverty than the less vulnerable tracts.}
    
    \begin{tabular}{l|ll|ll|ll}
    
        \toprule
        \multicolumn{7}{c}{\textbf{\% Households in Poverty}} \\
        \midrule
        {} & \multicolumn{2}{l}{Mean} & \multicolumn{2}{l}{Median} & \multicolumn{2}{l}{Std Dev} \\
        {Housing Vulnerability} & \textit{Less}  & \textit{Most}  & \textit{Less}  & \textit{Most}  & \textit{Less}  & \textit{Most}  \\
        City         &            &       &        &       &         &       \\
        \midrule
        Albuquerque  &      10.17 & 16.52 &   7.64 & 15.06 &    6.55 &  9.77 \\
        Boston       &      11.47 & 18.00 &   8.49 & 14.77 &    9.13 & 12.31 \\
        Bridgeport   &       5.02 & 19.51 &   3.80 & 17.35 &    4.42 & 13.28 \\
        Charleston   &       9.01 & 21.36 &   6.84 & 20.41 &    6.96 & 12.83 \\
        Cincinnati   &       7.25 & 28.19 &   5.97 & 25.42 &    6.68 & 16.27 \\
        Cleveland    &       7.91 & 34.79 &   6.10 & 33.39 &    5.84 & 11.98 \\
        Columbus     &      11.08 & 30.05 &   9.02 & 27.56 &    8.00 & 12.56 \\
        Dallas       &      10.05 & 15.53 &   7.33 & 14.28 &    9.45 & 10.58 \\
        Fort Worth   &       6.21 & 16.60 &   4.82 & 14.38 &    5.30 & 11.54 \\
        Gainesville  &      15.75 & 28.69 &   6.47 & 27.44 &   20.94 & 16.27 \\
        Greenville   &       7.72 & 19.53 &   4.58 & 14.35 &    8.76 & 12.11 \\
        Hartford     &       6.79 & 29.29 &   6.13 & 27.42 &    4.06 & 11.40 \\
        Houston      &       7.07 & 19.50 &   5.03 & 17.23 &    7.03 & 12.92 \\
        Jacksonville &      10.01 & 24.65 &   7.71 & 23.65 &    6.71 & 12.66 \\
        Kansas City  &       9.21 & 23.94 &   6.93 & 24.19 &    8.27 & 10.38 \\
        Las Vegas    &       7.15 & 21.53 &   6.39 & 22.55 &    4.13 & 11.10 \\
        Milwaukee    &       9.23 & 33.98 &   7.74 & 33.16 &    6.76 & 14.10 \\
        New Orleans  &      16.48 & 30.44 &  13.33 & 28.34 &   11.86 & 13.97 \\
        Philadelphia &      12.60 & 34.08 &  11.46 & 32.61 &    8.15 & 12.96 \\
        St Louis     &      12.54 & 29.32 &  10.69 & 28.25 &    6.74 &  8.06 \\
        \bottomrule
        \end{tabular}
    \label{tab:ses_poverty}
\end{table}

\begin{table}[]
    \caption{Statistical properties of transit-based commuting levels in the census tracts that are less and most vulnerable to housing insecurity. The most vulnerable tracts have higher rates of transit-based commutes than the less vulnerable tracts.}

    \centering
    \begin{tabular}{l|ll|ll|ll}
        \toprule
        \multicolumn{7}{c}{\textbf{\% Commuting via Public Transit}} \\
        \midrule
        {} & \multicolumn{2}{l}{Mean} & \multicolumn{2}{l}{Median} & \multicolumn{2}{l}{Std Dev} \\
        {Housing Vulnerability} & \textit{Less}  & \textit{Most}  & \textit{Less}  & \textit{Most}  & \textit{Less}  & \textit{Most}  \\
        City         &                  &       &        &       &         &       \\
        \midrule
        Albuquerque  &             0.98 &  1.19 &   0.54 &  0.67 &    1.23 &  1.56 \\
        Boston       &            27.11 & 32.43 &  27.57 & 28.56 &   10.54 & 17.71 \\
        Bridgeport   &             7.89 & 10.69 &   6.86 &  9.15 &    6.31 &  7.02 \\
        Charleston   &             0.71 &  3.42 &   0.00 &  1.20 &    1.43 &  5.91 \\
        Cincinnati   &             1.51 &  8.84 &   0.77 &  6.74 &    2.54 &  7.50 \\
        Cleveland    &             2.07 & 13.16 &   1.22 & 10.05 &    2.93 &  9.29 \\
        Columbus     &             1.30 &  5.55 &   0.27 &  4.79 &    2.54 &  4.58 \\
        Dallas       &             1.60 &  1.99 &   0.77 &  1.09 &    2.80 &  2.78 \\
        Fort Worth   &             0.41 &  0.60 &   0.00 &  0.00 &    0.85 &  1.20 \\
        Gainesville  &             1.76 &  5.82 &   0.13 &  3.47 &    2.51 &  6.62 \\
        Greenville   &             0.14 &  1.10 &   0.00 &  0.00 &    0.36 &  1.74 \\
        Hartford     &             1.27 & 11.54 &   0.80 &  8.25 &    1.45 &  9.99 \\
        Houston      &             1.48 &  2.68 &   0.84 &  1.16 &    1.95 &  3.87 \\
        Jacksonville &             0.96 &  4.51 &   0.25 &  2.95 &    1.70 &  5.16 \\
        Kansas City  &             0.60 &  6.38 &   0.00 &  4.52 &    1.43 &  6.62 \\
        Las Vegas    &             1.16 &  7.13 &   0.50 &  5.43 &    1.72 &  6.64 \\
        Milwaukee    &             2.73 & 10.33 &   1.79 &  7.89 &    3.31 &  8.56 \\
        New Orleans  &             5.05 &  9.95 &   4.41 &  8.41 &    4.04 &  8.55 \\
        Philadelphia &            20.50 & 33.42 &  18.85 & 33.03 &   10.91 & 11.59 \\
        St Louis     &             3.65 & 16.04 &   3.78 & 16.23 &    2.54 & 10.15 \\
        \bottomrule
        \end{tabular}
    \label{tab:ses_ptCommuteMode}
\end{table}

\begin{table}[]
    \centering
    \caption{Statistical properties of commuting times in the census tracts that are less and most vulnerable to housing insecurity. The distinction between housing demographics based on commuting times tends to be less clear than other sociodemographic indicators.}
    \begin{tabular}{l|ll|ll|ll}
        \toprule
        \multicolumn{7}{c}{\textbf{\% Commuting over an Hour}} \\
        \midrule
        {} & \multicolumn{2}{l}{Mean} & \multicolumn{2}{l}{Median} & \multicolumn{2}{l}{Std Dev} \\
        {Housing Vulnerability} & \textit{Less}  & \textit{Most}  & \textit{Less}  & \textit{Most}  & \textit{Less}  & \textit{Most}  \\
        City         &                &       &        &       &         &      \\
        \midrule
        Albuquerque  &           5.35 &  3.90 &   4.81 &  2.62 &    3.90 & 4.59 \\
        Boston       &          10.51 &  8.47 &   9.28 &  7.01 &    5.84 & 6.53 \\
        Bridgeport   &          16.88 & 13.04 &  15.53 & 12.19 &    7.15 & 5.94 \\
        Charleston   &           3.20 &  4.98 &   1.81 &  3.72 &    4.00 & 5.59 \\
        Cincinnati   &           3.26 &  5.89 &   2.49 &  4.82 &    2.77 & 4.62 \\
        Cleveland    &           4.05 &  7.29 &   3.33 &  5.94 &    3.47 & 5.84 \\
        Columbus     &           3.86 &  5.53 &   3.21 &  4.85 &    3.42 & 3.96 \\
        Dallas       &           6.96 &  8.49 &   6.19 &  7.46 &    5.32 & 6.53 \\
        Fort Worth   &           6.97 &  7.89 &   6.25 &  7.25 &    4.52 & 4.35 \\
        Gainesville  &           4.05 &  3.48 &   3.55 &  2.95 &    2.98 & 2.73 \\
        Greenville   &           3.72 &  4.48 &   3.34 &  3.71 &    2.78 & 3.64 \\
        Hartford     &           4.10 &  5.58 &   3.81 &  4.67 &    2.33 & 3.99 \\
        Houston      &          10.52 & 10.23 &  10.00 &  8.77 &    7.02 & 6.58 \\
        Jacksonville &           4.33 &  5.61 &   4.08 &  4.10 &    3.11 & 3.86 \\
        Kansas City  &           3.92 &  3.70 &   2.85 &  3.60 &    3.82 & 3.42 \\
        Las Vegas    &           3.49 &  6.21 &   2.75 &  5.31 &    3.12 & 4.56 \\
        Milwaukee    &           3.38 &  5.51 &   2.87 &  4.49 &    2.68 & 4.12 \\
        New Orleans  &           5.54 &  6.96 &   4.29 &  5.77 &    4.69 & 5.30 \\
        Philadelphia &          14.60 & 15.79 &  13.11 & 14.47 &    7.73 & 7.45 \\
        St Louis     &           3.45 &  9.84 &   2.84 &  7.97 &    1.99 & 7.41 \\
        \bottomrule
        \end{tabular}
    \label{tab:ses_ptCommuteTime}
\end{table}

\section{Public Transportation}

\begin{table}[]

    \caption{Statistical properties of transit efficiency and area ($km^2$) for 20 US cities, ranked by decreasing efficiency.}
    \centering
    \begin{tabular}{l|rrr|r}
\toprule
\multicolumn{1}{l}{City} & \multicolumn{3}{l}{Transit Efficiency} & \multicolumn{1}{l}{Area ($km^2$)}\\
& \textit{Mean} & \textit{Median} & \textit{Std. Dev.}    &           \\
\midrule
Philadelphia  &  0.914 &   0.828 & 0.357 &   347.782 \\
Milwaukee     &  1.314 &   1.145 & 0.631 &   625.435 \\
Boston        &  1.839 &   1.744 & 0.690 &   150.863 \\
San Francisco &  1.975 &   1.888 & 0.534 &   121.478 \\
Cleveland     &  2.262 &   2.061 & 0.816 &  1184.094 \\
Cincinnati    &  2.277 &   1.979 & 1.003 &  1050.000 \\
New Orleans   &  2.842 &   2.720 & 0.964 &   438.831 \\
Jacksonville  &  3.081 &   2.626 & 1.613 &  1975.201 \\
Albuquerque   &  3.494 &   3.001 & 1.582 &  3007.645 \\
Las Vegas     &  3.653 &   2.989 & 2.770 & 20439.277 \\
Houston       &  4.068 &   3.454 & 2.042 &  9350.383 \\
Columbus      &  4.250 &   3.282 & 2.352 &  3833.131 \\
Charleston    &  4.397 &   3.868 & 2.304 &  2377.463 \\
Gainesville   &  4.542 &   3.917 & 2.475 &  2267.635 \\
Kansas City   &  4.547 &   3.735 & 2.836 &  5487.102 \\
Greenville    &  5.073 &   4.710 & 1.985 &  2033.481 \\
Fort Worth    &  5.994 &   5.883 & 2.297 &  2236.864 \\
Bridgeport    &  6.315 &   6.277 & 2.845 &  1618.659 \\
Hartford      &  8.420 &   8.561 & 2.231 &  1903.544 \\
Dallas        & 10.573 &  10.744 & 2.505 &  2259.440 \\
\bottomrule
\end{tabular}
    \label{si:transit_efficiency}
\end{table}

\begin{figure}[]
  \centering
  \includegraphics[width=1 \textwidth]{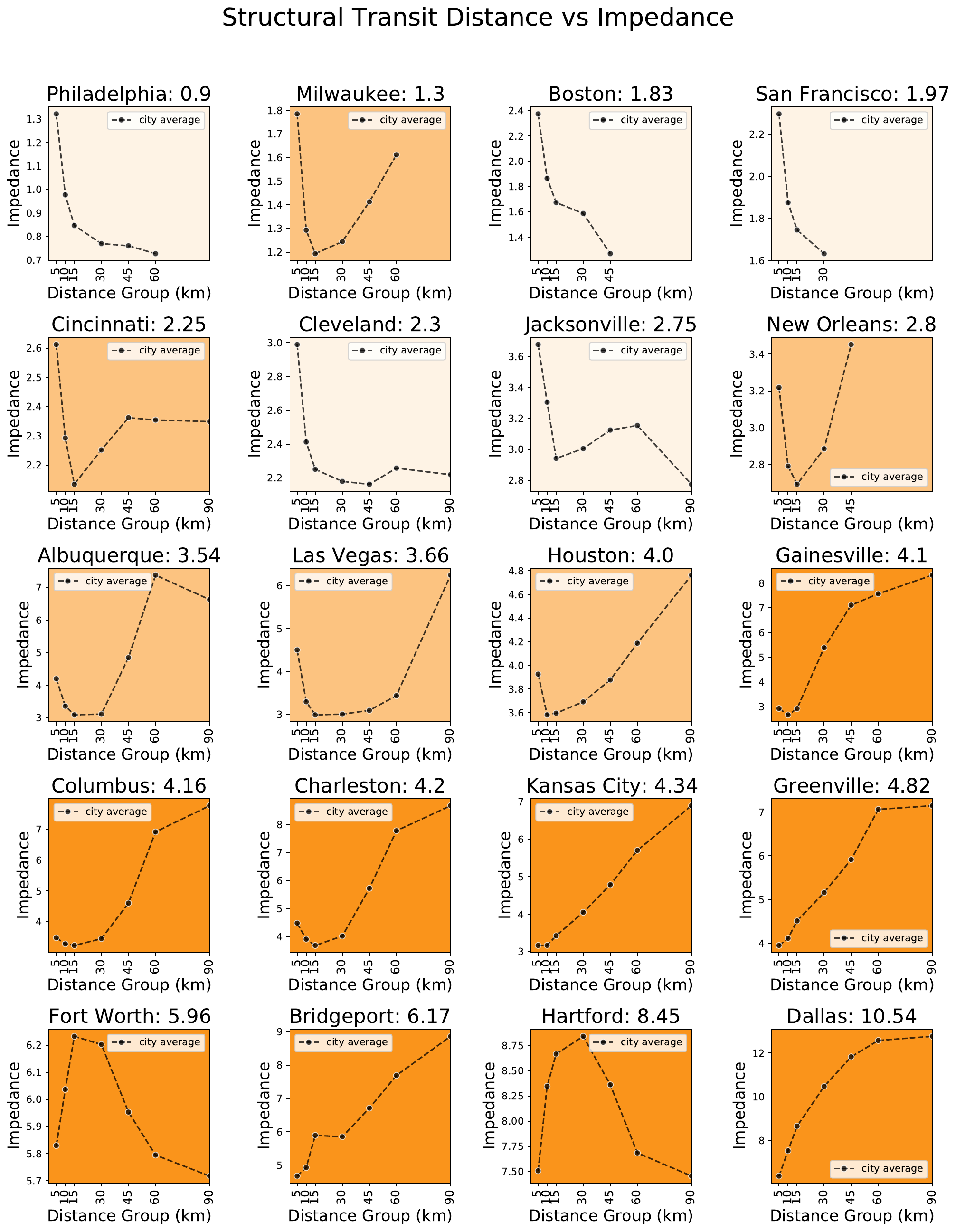}
  \caption{Transit efficiency as a function of trip distance for 20 US cities. Lighter orange plots reflect efficient systems in which travel impedance decreases as trip distance increases (Philadelphia, Boston, San Francisco, Cleveland, Jacksonville. Dark orange plots indicate inefficient transit systems, with increasing travel impedance as trip distance increases (Gainesville, Columbus, Charleston, Kansas City, Greenville, Fort Worth, Bridgeport, Hartford, Dallas). The remaining orange plots capture cities with moderately efficient transit service, in which the relationship between travel impedance and trip distance switches from negative to positive at a given trip distance threshold. }
  \label{fig:allCities_transitImpVStripDist}
  
\end{figure}

\subsection{Exploring the Intersection of the Housing and Transit Landscape\label{ssec:static_housing_and_transit}}

\begin{figure}[H]
  \centering
  \includegraphics[width=0.8\textwidth]{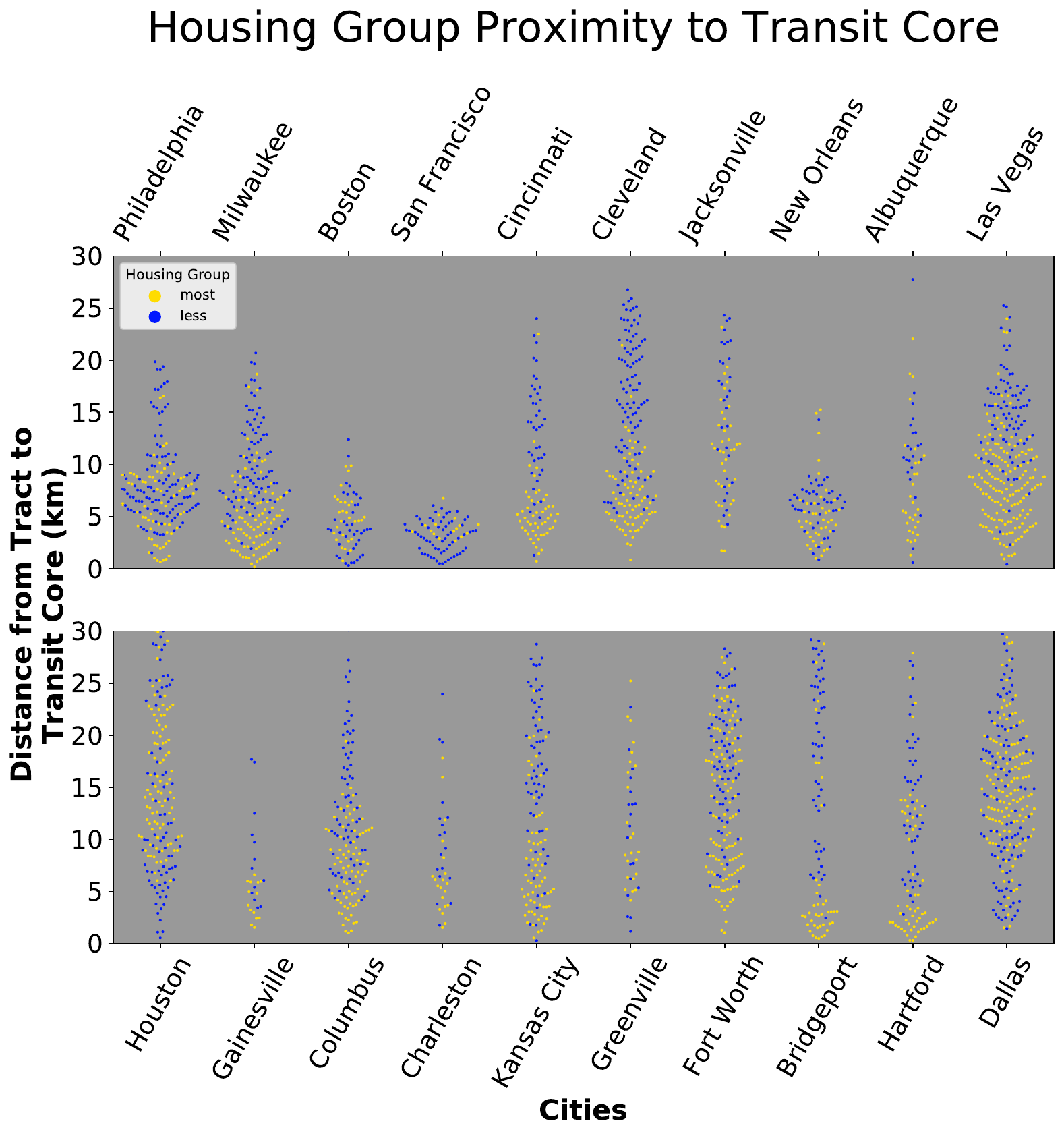}
  \caption{Swarmplot that examines differences between housing demographics, based on their proximity to the transit core. The y-axis is the distance in kilometres, while the x-axis reflects the distribution for each city. Blue dots represent the less vulnerable housing demographic while yellow symbolise tracts that are most vulnerable to housing insecurity.}
  \label{fig:tract_prox_to_core}

\end{figure}

Prior to understanding the role transit plays in constraining or facilitating mobility for various housing groups, we analyse how housing demographics are spatially distributed around the public transportation infrastructure. Specifically, we consider how close each tract is to the transit system's center of mass, which we refer to as the \textit{transit\ core}. We use the center of mass as a proxy for the location that has the most access to different transit stops in the region. We define the transit core by calculating the average longitude and latitude for all transit stops in the system, weighting for frequency of trips through each transit stop. Figure \ref{fig:tract_prox_to_core} illustrates the disparity in residential locations for the most and less vulnerable housing groups. That is, in the vast majority of cities, the tracts that are most vulnerable to housing insecurity tend to be situated closer to the \textit{transit core}, with the exceptions being Philadelphia, Boston, San Francisco, Houston, and Dallas. Cities are ordered in increasing transit efficiency, with Philadelphia having the most efficient transportation system and Dallas being characterised by the least efficient transit infrastructure. The mean distances to the transit core for each city's housing demographic is listed in Table \ref{si:core_dist} in the Supplementary Materials. We can observe that the cities with more efficient transit networks, such as Philadelphia, Boston, and Milwaukee, express a less distinct separation in proximity of each housing group.

This analysis paints a clearer picture of how cities with less robust transit can increase transport inequality by forcing dependence on individual vehicles. Furthermore, cities with inefficient transit systems provide less efficient service to areas further away, which could perpetuate levels of residential segregation. We observe how the most vulnerable housing demographics tend to live closer to the transit core than their less vulnerable counterparts. This trend is consistent with transportation poverty studies that identify less privileged residents living closer to urban infrastructure and services \cite{allen2019sizing}. These findings emphasise the importance of space in the housing landscape, as urban constraints may force vulnerable populations to live in the core of a city. Consequently, these demographic groups will appear to have better access to resource, yet the lack of choice is obfuscated. The next section focuses on exploring whether the identified difference is a constraint or an advantage to populations that are most vulnerable to housing insecurity.

\begin{table}
\centering

\caption{\label{tract_mean_prox_to_core}Mean distance from tracts to transit core}
\begin{tabular}{lccc}
\toprule
 &   Most &   Mildly &   Less \\ 
 City       & Vulnerable & Vulnerable & Vulnerable \\
\midrule
Philadelphia  &  6.267 &  6.729 &  8.512 \\
Milwaukee     &  5.003 &  6.889 &  9.650 \\
Boston        &  5.007 &  5.571 &  4.085 \\
San Francisco &  3.959 &  4.454 &  3.065 \\
Cincinnati    &  5.355 &  9.800 & 13.344 \\
Cleveland     &  7.063 & 11.800 & 16.417 \\
Jacksonville  & 10.702 & 10.226 & 15.331 \\
New Orleans   &  5.193 &  5.218 &  6.116 \\
Albuquerque   &  8.530 &  9.107 & 11.138 \\
Las Vegas     &  8.333 & 18.854 & 17.406 \\
Houston       & 19.868 & 22.007 & 21.452 \\
Gainesville   &  4.160 & 10.509 &  8.510 \\
Columbus      &  8.514 & 13.957 & 17.122 \\
Charleston    &  8.338 & 12.028 & 10.397 \\
Kansas City   &  8.594 & 17.018 & 22.131 \\
Greenville    & 13.575 &  8.748 & 10.541 \\
Fort Worth    & 13.530 & 15.031 & 17.993 \\
Bridgeport    & 11.411 & 18.355 & 22.971 \\
Hartford      &  6.749 & 14.600 & 13.577 \\
Dallas        & 15.034 & 14.182 & 13.770 \\
\bottomrule
\end{tabular}
\label{si:core_dist}
\end{table}

\section{Spatial Autocorrelation}

Table \ref{tab:gsa} lists the Moran's I statistic used to test global spatial autocorrelation in workplaces for each of the housing demographics. This is performed by using the fraction of a census tract's workforce that is made to examine whether there is a spatial relationship between these employment rates.
\begin{table}[htbp]
    \caption{Moran's I statistic for census tracts based on the fraction of individuals from each housing demographic that work there.}\label{tab:gsa}
    \centering
    \setlength\tabcolsep{12pt}
    \begin{tabular}{@{}llll@{}}
    \toprule%
    \multirow{2}{*}[3pt]{City}
      &  \multicolumn{3}{c}{Moran's \textit{I} for Housing Demographics' Workplaces} \\
      \cmidrule(lr){2-4}
    & Less &               Mild &               Most \\
    \midrule
    Gainesville   &   \textbf{0.219**} &              0.094 &    \textbf{0.115*} \\
    New Orleans   &  \textbf{0.192***} &              0.014 &   \textbf{0.132**} \\
    Greenville    &  \textbf{0.250***} &  \textbf{0.516***} &  \textbf{0.345***} \\
    Fort Worth    &  \textbf{0.381***} &  \textbf{0.388***} &  \textbf{0.393***} \\
    Dallas        &  \textbf{0.329***} &  \textbf{0.303***} &  \textbf{0.406***} \\
    Boston        &  \textbf{0.348***} &  \textbf{0.540***} &  \textbf{0.435***} \\
    Las Vegas     &  \textbf{0.429***} &  \textbf{0.253***} &  \textbf{0.437***} \\
    Charleston    &  \textbf{0.368***} &              0.052 &  \textbf{0.453***} \\
    San Francisco &  \textbf{0.543***} &  \textbf{0.654***} &  \textbf{0.499***} \\
    Albuquerque   &   \textbf{0.203**} &  \textbf{0.273***} &  \textbf{0.506***} \\
    Cleveland     &  \textbf{0.587***} &  \textbf{0.249***} &  \textbf{0.552***} \\
    Hartford      &  \textbf{0.287***} &  \textbf{0.385***} &  \textbf{0.558***} \\
    Cincinnati    &  \textbf{0.424***} &  \textbf{0.582***} &  \textbf{0.560***} \\
    Columbus      &  \textbf{0.467***} &  \textbf{0.644***} &  \textbf{0.615***} \\
    New York City &  \textbf{0.709***} &  \textbf{0.645***} &  \textbf{0.621***} \\
    Jacksonville  &  \textbf{0.352***} &  \textbf{0.512***} &  \textbf{0.631***} \\
    Milwaukee     &  \textbf{0.622***} &  \textbf{0.309***} &  \textbf{0.638***} \\
    Houston       &  \textbf{0.430***} &  \textbf{0.562***} &  \textbf{0.640***} \\
    Philadelphia  &  \textbf{0.620***} &  \textbf{0.559***} &  \textbf{0.656***} \\
    St Louis      &  \textbf{0.650***} &  \textbf{0.423***} &  \textbf{0.678***} \\
    Bridgeport    &  \textbf{0.483***} &  \textbf{0.205***} &  \textbf{0.720***} \\
    Kansas City   &  \textbf{0.630***} &  \textbf{0.523***} &  \textbf{0.767***} \\
    \bottomrule
    \end{tabular}
    \footnotetext[1]{\text{*p}$< 0.05; $\text{**p} $<0.01; $ \text{***p}$<0.001$}
    \label{tab:gsa}

\end{table}

A higher coefficient implies that tracts that hire a large fraction of employees from a particular housing demographic tend to be near other tracts that exhibit a similar behaviour. This also implies that areas that hire a lower rate of individuals from a housing demographic tend to be situated around other tracts that employ a similar rate of workers from that demographic. Table \ref{tab:gsa} shows how most cities have a spatial component to employment behaviour of the most vulnerable demographic, with space playing a smaller role in Gainesville and New Orleans.
\bibliography{sn-article}
